\documentclass[a4paper,11pt,pdfusetitle]{article}
\usepackage{jheppub}

\usepackage[utf8]{inputenc}
\usepackage{amsmath}
\usepackage{amssymb}
\usepackage{pifont}
\usepackage[dvipsnames]{xcolor}
\usepackage{graphicx}
\usepackage{dcolumn}
\usepackage{bm}
\usepackage{physics}
\usepackage{rotating}
\usepackage{pdflscape}

\hypersetup{allcolors=[rgb]{1,0.56,0}}

\newcommand{\orcid}[1]{\href{https://orcid.org/#1}{#1}}
\newcommand{\dPDG}{\delta_{\rm PDG}}
\newcommand{\e}[1]{\times10^{#1}}

\title{The Impact of Different Parameterizations on the Interpretation of CP Violation in Neutrino Oscillations}

\author[a,1]{Peter B.~Denton\note{\orcid{0000-0002-5209-872X}}}
\author[a,b,2]{and Rebekah Pestes\note{\orcid{0000-0002-9634-1664}}}

\affiliation[a]{High Energy Theory Group, Physics Department, Brookhaven National Laboratory, Upton, NY 11973, USA}
\affiliation[b]{Center for Neutrino Physics, Department of Physics, Virginia Tech, Blacksburg, VA 24061, USA}

\emailAdd{pdenton@bnl.gov}
\emailAdd{rebhawk8@vt.edu}

\date{January 2, 2021}

\abstract{
CP violation in the lepton mass matrix will be probed with good precision in upcoming experiments.
The amount of CP violation present in oscillations can be quantified in numerous ways and is typically parameterized by the complex phase $\delta_{\rm PDG}$ in the standard PDG definition of the lepton mixing matrix.
There are additional parameterizations of the lepton mixing matrix as well.
Through various examples, we explore how, given the current data, different parameterizations can lead to different conclusions when working with parameterization dependent variables, such as $\delta$.
We demonstrate how the smallness of $|U_{e3}|$ governs the scale of these results.
We then demonstrate how $\delta$ can be misleading and argue that the Jarlskog is the cleanest means of presenting the amount of CP violation in the lepton sector.
We also confirm that, among the different parameterizations considered, the standard PDG parameterization has a number of convenient features.}

\begin{document}

\maketitle

\section{Introduction}
While the fact that CP violation (CPV) is needed to explain the matter-antimatter asymmetry in the universe \cite{Sakharov:1967dj} is a frequently used motivation for searching for sources of CPV, identifying sources of CPV is interesting in its own right.
Not only does it play an important part of the ongoing goal of measuring the parameters of the Standard Model, but also for understanding when CP is and is not violated.
In the quark sector, the Cabibo-Kobayashi-Maskawa (CKM) matrix \cite{Cabibbo:1963yz,Kobayashi:1973fv} provides a small amount of CPV \cite{Tanabashi:2018oca}.
For leptons, three phases relating to how neutrinos and anti-neutrinos behave differently appear in the lepton mixing matrix $U$ \cite{Pontecorvo:1957cp,Maki:1962mu}, one of which ($\delta$) is analogous to the KM phase in the CKM matrix while the other two are physical if and only if neutrinos have a Majorana contribution to their mass.
$U$, like the CKM matrix for quarks, describes the mismatch of lepton flavor eigenstates ($\ket{\nu_\alpha}$ for $\alpha=e,\mu,\tau$) and mass eigenstates ($\ket{\nu_i}$ for $i=1,2,3$) via $\ket{\nu_\alpha}=\sum_iU_{\alpha i}^*\ket{\nu_i}$.
While all three phases contribute to leptonic CPV, the complex phase $\delta$ contributes to neutrino oscillation while the two Majorana phases, though important for neutrinoless double beta decay, do not. The complex phase $\delta$, however, is convention dependent and thus not a fundamental quantity.
The most useful fundamental quantity to describe the amount of CPV in a mass matrix is the Jarlskog invariant \cite{Jarlskog:1985ht}:
\begin{equation}
J\equiv\Im[U_{e1}U_{e2}^*U_{\mu1}^*U_{\mu2}]\,.
\label{eq:jarlskog}
\end{equation}
The choice of the $2\times2$ submatrix in eq.~\ref{eq:jarlskog} is arbitrary; we chose the (3,$\tau$) submatrix resulting from deleting the third column and the $\tau$ row in eq.~\ref{eq:jarlskog} for concreteness.
Any row and any column can be removed.

There are many different ways of parameterizing the lepton mixing matrix, all of which must have at least one complex phase.
Depending on the details of the parameterization, the constraint on the complex phase can be quite different.
This means that interpreting results and goals of experiments in terms of the complex phase of one parameterization is not a fundamental description of our understanding of CPV.

In addition, new physics scenarios, if they exist in reality, can fundamentally complicate the extraction of CP violation independent of how the mixing matrix is parameterized.
For example, the addition of non-standard neutrino interactions (with or without additional CP violating phases), sterile neutrinos (also with or without additional CP violating phases), or unitary violation\footnote{A general non-unitary $3\times3$ complex matrix has 18 free parameters; for neutrino oscillations five of the complex phases can be removed via rephasing.
While one has a choice in which row and column to make real (or any other rephasing choice) this only leads to a constant shift in the complex phases.} can all lead to confusion related to the extraction of the standard CP violating three-flavor phase \cite{GonzalezGarcia:2001mp,Coloma:2011rq,Friedland:2012tq,Gandhi:2015xza,Rahman:2015vqa,Masud:2015xva,Palazzo:2015gja,Coloma:2015kiu,deGouvea:2015ndi,Masud:2016bvp,Agarwalla:2016xxa,Dutta:2016glq,Liao:2016hsa,Liao:2016orc,Ge:2016dlx,Fukasawa:2016lew,Forero:2016cmb,deGouvea:2016pom,Miranda:2016wdr,Choubey:2017cba,Deepthi:2017gxg,Hyde:2018tqt,Gupta:2018qsv,Branco:2019avf,Denton:2020uda}.
Some of these degeneracies can be differentiated by combining measurements from different experiments.
These scenarios are different from those discussed in this paper in that in the presence of new physics, extracting information about CP violation would suffer from various partial degeneracies \emph{independent} of the choice of parameterization.
The optimal choice of parameterization of the $3\times3$ (or $4\times4$ or larger matrix in the presence of additional neutrinos) mixing matrix in the presence of various new physics scenarios is beyond the scope of this article and we will focus on the standard three-flavor oscillation picture with no new interactions.

In this paper, we will demonstrate exactly how the complex phase of the lepton mixing matrix is not the optimal parameter for understanding CPV in neutrino oscillations and when it can be misleading.
We will show how different, perfectly valid, representations of the mixing matrix with one complex phase can lead to rather different conclusions when the complex phase is used as the primary indicator of CPV in oscillations.
In section \ref{sec:parameterizations}, we will define the parameterizations we are going to consider in this paper.
Next, we will show how these different parameterizations affect their respective $\delta'$ and the key role that $U_{e3}$ plays in section \ref{sec:comparison}.
Finally, we will discuss our results in the context of T2K's measurements in section \ref{sec:discussion} and we will draw several interesting conclusions in section \ref{sec:conclusion}.
We also discuss several additional scenarios in the appendices including unitary violation, the quark mixing matrix, and flavor models.

\section{Mixing Matrix Parameterizations}
\label{sec:parameterizations}
We anticipate that the lepton mixing matrix, $U$, should be unitary up to any corrections from sterile neutrinos, which we ignore (for a discussion of parameterizations and general unitary violation schemes, see appendix \ref{sec:uv}).
Thus, the matrix can be parameterized as three rotation matrices, containing a total of three Euler mixing angles ($\theta_{12},\theta_{13},\theta_{23}$) and six complex phases.
However, all rows and columns can be individually rephased if neutrinos are Dirac, or if we restrict ourselves to relativistic neutrinos, as is the case for neutrino oscillations.
Rephasing is the fact that any column or row can be multiplied by an arbitrary phase.
While this appears to remove all six phases present in an arbitrary $3\times3$ unitary matrix, one of these six rephasings is dependent on the other five, which leaves one complex phase remaining for describing neutrino oscillations and is usually labeled as $\delta$.
Rephasing invariance also allows us to constrain each mixing angle to a quadrant of our choice, since changing the quadrant of a mixing angle is equivalent to specific combinations of rephasing and shifting $\delta$ by a constant.
We choose to constrain all of the mixing angles to the first quadrant ($\theta_{12},\theta_{13},\theta_{23}\in[0,90^\circ)$).
The three rotation matrices can be in any order, and the complex phase $\delta$ can be put into any of the rotation matrices.

When we make various parameterizations of $U$ for neutrino oscillations, the magnitudes of the matrix elements remain the same, as does the Jarlskog, but the four parameters change; this is the crux of our paper.
Changing the order of the rotation matrices changes the values of the mixing angles and $\delta$ (see table \ref{tab:parameters}), but changing which rotation $\delta$ is on does not (other than trivial redefinitions, such as $\delta\to-\delta$), so we will only consider having $\delta$ in the rotation matrix containing $\theta_{13}$.
Thus, the rotation matrices are defined as follows:
\begin{equation}
U_{1} \equiv 
\begin{pmatrix}
1 & 0 & 0 \\ 0 & c_{23} & s_{23} \\ 0 & -s_{23} & c_{23}
\end{pmatrix}
\,\text{,}\hspace{5mm}
U_{2} \equiv 
\begin{pmatrix}
c_{13} & 0 & s_{13}e^{-i\delta} \\ 0 & 1 & 0 \\ -s_{13}e^{i\delta} & 0 & c_{13} 
\end{pmatrix}
\,\text{, }\hspace{3mm}\text{ and }\hspace{3mm}
U_{3} \equiv 
\begin{pmatrix}
c_{12} & s_{12} & 0 \\ -s_{12} & c_{12} & 0 \\ 0 & 0 & 1
\end{pmatrix}
\,\text{, }
\label{eq:U defn}
\end{equation}
where $c_{ij}\equiv\cos(\theta_{ij})$ and $s_{ij}\equiv\sin(\theta_{ij})$.
In this paper, we are only considering parameterizations of the form $U_{ijk}\equiv U_i U_j U_k$ for $i,j,k=1,2,3$ in which $i$, $j$, and $k$ are all different.
Other parameterizations exist \cite{Schechter:1980gr,Gerard:2012ft} including those using the same rotation axis twice ($U_{iji}$) \cite{Fritzsch:2001ty}, three of the Gell-Mann matrices parameterizing the generators of SU(3) \cite{Merfeld:2014cha,Boriero:2017tkh,Davydova:2019aat,Zhukovsky:2019jzz}, four complex phases \cite{Aleksan:1994if}, the exponential of a complex matrix \cite{Zhukovsky:2016mon}, or five rotations and a complex phase \cite{Emmanuel-Costa:2015tca}\footnote{The primary parameterization presented in \cite{Emmanuel-Costa:2015tca} has the same complex phase as in the $U_{132}$ parameterization below.}.
If one wanted to include the Majorana phases in the parameterization for $U$, the symmetric parameterization \cite{Schechter:1980gk,Rodejohann:2011vc} is advantageous over the PDG method of including the Majorana phases, and it simplifies to the PDG parameterization in the context neutrino oscillations.
In this article, we focus on parameterizations containing three rotations and one complex phase, but we also show numerical results for the case of repeated rotations of the form $U_{iji}$ confirming that they are qualitatively the same as those of the form $U_{ijk}$.
In addition, we discuss the CKM matrix in different parameterizations in appendix \ref{sec:ckm}.

We also note that there are numerous other models of neutrino mixing designed to predict certain relationships among the oscillation parameters, see e.g.~\cite{Altarelli:2010gt,King:2013eh,Branco:2014zza,Petcov:2017ggy,Xing:2019vks,Smirnov:2018luj,Smirnov:2019msn,Gehrlein:2020jnr}.
As these models have fewer than four parameters, they don't span the entire relevant space and thus any features related to the complex phase would depend on correlations with the other parameters introduced by the model choices.
For this reason, along with the very large body of literature on this topic, we do not present a comprehensive review of the impact of parameterization in neutrino flavor models, but we encourage the choice of parameterization to be carefully examined as these models are developed.

The parameterization of $U$ used by the Particle Data Group (PDG) \cite{Tanabashi:2018oca} without the Majorana phases is
\begin{equation}
U_\text{PDG} = U_{123} =
\begin{pmatrix}
c_{12}c_{13}
& s_{12}c_{13}
& s_{13}e^{-i\delta} \\
-s_{12}c_{23}-c_{12}s_{13}s_{23}e^{i\delta}
& c_{12}c_{23}-s_{12}s_{13}s_{23}e^{i\delta}
& c_{13}s_{23} \\
s_{12}s_{23}-c_{12}s_{13}c_{23}e^{i\delta}
& -c_{12}s_{23}-s_{12}s_{13}c_{23}e^{i\delta}
& c_{13}c_{23}
\end{pmatrix}\,,
\label{eq:UPDG}
\end{equation}
and the Jarlskog in this parameterization is
\begin{equation}
J=c_{12}s_{12}c_{13}^2s_{13}c_{23}s_{23}\sin(\dPDG)\,.
\end{equation}
The mixing matrix and Jarlskog in the other five parameterizations considered are shown in table \ref{tab:matJ}, see also ref.~\cite{Rasin:1997pn} for a related discussion for the quark mixing matrix.

We note some important features of this parameterization.
The most important of which is the difference between ``simple'' elements and ``complicated'' elements.
We define a simple element to be one that is only a product of trigonometric functions and $e^{\pm i\delta}$ terms while complicated functions can also be the sum or difference of such terms.
Thus in $U_{123}$, we see that the first row and third column are composed entirely of simple elements.
In general, using the definitions in eq.~\ref{eq:U defn}, $U_{ijk}$ has simple elements along the $i^{\rm th}$ row and the $k^{\rm th}$ column.

In the PDG convention, the complex phase $\delta$ is set such that $\arg(U_{e3})=-\delta$.
Within the PDG convention, this phase can be shifted with no effect on any of the oscillation physics.
For example, one could multiply the third column by $e^{i\delta}$ and then the second and third rows by $e^{-i\delta}$ so that the first row and third column were all real.

\begin{landscape}
\begin{table}
    \centering
    \caption{Parameterizations of the neutrino mixing matrix $U$ under consideration in this paper and their corresponding Jarlskog invariant. The primed variables ($\delta'$, $\theta_{ij}'$, $c_{ij'}\equiv\cos(\theta_{ij}')$, $s_{ij'}\equiv\sin(\theta_{ij}')$) denote parameters in the specified alternative parameterization, whereas unprimed variables ($\dPDG$, $\theta_{ij}$, $c_{ij}\equiv\cos(\theta_{ij})$, $s_{ij}\equiv\sin(\theta_{ij})$) denote parameters in $U_\text{PDG}$.}
    \label{tab:parameterizations}
    \begin{tabular}{|l||c|c|}
        \hline
        $U_{ijk}$ & Matrix & $J$ \\
        \hline\hline
        $U_{123}$
            & $\begin{pmatrix}
                c_{12}c_{13}
                    & s_{12}c_{13}
                    & s_{13}e^{-i\delta} \\
                -s_{12}c_{23}-c_{12}s_{13}s_{23}e^{i\delta}
                    & c_{12}c_{23}-s_{12}s_{13}s_{23}e^{i\delta}
                    & c_{13}s_{23} \\
                s_{12}s_{23}-c_{12}s_{13}c_{23}e^{i\delta}
                    & -c_{12}s_{23}-s_{12}s_{13}c_{23}e^{i\delta}
                    & c_{13}c_{23}
                \end{pmatrix}$
            & $c_{12}s_{12}c_{13}^2s_{13}c_{23}s_{23}\sin(\dPDG)$
            \\
        \hline
        $U_{132}$
            & $\begin{pmatrix}
                c_{12'}c_{13'}
                    & s_{12'}
                    & c_{12'}s_{13'}e^{-i\delta'} \\
                -s_{12'}c_{13'}c_{23'}-s_{13'}s_{23'}e^{i\delta'}
                    & c_{12'}c_{23'}
                    & c_{13'}s_{23'} -s_{12'}s_{13'}c_{23'}e^{-i\delta'} \\
                s_{12'}c_{13'}s_{23'}-s_{13'}c_{23'}e^{i\delta'}
                    & -c_{12'}s_{23'}
                    & c_{12'}c_{23'}+s_{12'}s_{13'}s_{23'}e^{-i\delta'}
                \end{pmatrix}$
            & $c_{12'}^2s_{12'}c_{13'}s_{13'}c_{23'}s_{23'}\sin(\delta')$
            \\
        \hline
        $U_{213}$
            & $\begin{pmatrix}
                c_{12'}c_{13'}+s_{12'}s_{13'}s_{23'}e^{-i\delta'}
                    & s_{12'}c_{13'}-c_{12'}s_{13'}s_{23'}e^{-i\delta'}
                    & s_{13'}c_{23'}e^{-i\delta'} \\
                -s_{12'}c_{23'}
                    & c_{12'}c_{23'}
                    & s_{23'} \\
                s_{12'}c_{13'}s_{23'}-c_{12'}s_{13'}e^{i\delta'}
                    & -c_{12'}c_{13'}s_{23'}-s_{12'}s_{13'}e^{i\delta'}
                    & c_{13'}c_{23'}
                \end{pmatrix}$
            & $c_{12'}s_{12'}c_{13'}s_{13'}c_{23'}^2s_{23'}\sin(\delta')$
            \\
        \hline
        $U_{231}$
            & $\begin{pmatrix}
                c_{12'}c_{13'}
                    & s_{12'}c_{13'}c_{23'}-s_{13'}s_{23'}e^{-i\delta'}
                    & s_{12'}c_{13'}s_{23'}+s_{13'}c_{23'}e^{-i\delta'} \\
                -s_{12'}
                    & c_{12'}c_{23'}
                    & c_{12'}s_{23'} \\
                -c_{12'}s_{13'}e^{i\delta'}
                    & -c_{13'}s_{23'}-s_{12'}s_{13'}c_{23'}e^{i\delta'}
                    & c_{13'}c_{23'}-s_{12'}s_{13'}s_{23'}e^{i\delta'}
                \end{pmatrix}$
            & $c_{12'}^2s_{12'}c_{13'}s_{13'}c_{23'}s_{23'}\sin(\delta')$
            \\
        \hline
        $U_{312}$
            & $\begin{pmatrix}
                c_{12'}c_{13'}-s_{12'}s_{13'}s_{23'}e^{i\delta'}
                    & s_{12'}c_{23'}
                    & s_{12'}c_{13'}s_{23'}+c_{12'}s_{13'}e^{-i\delta'} \\
                -s_{12'}c_{13'}-c_{12'}s_{13'}s_{23'}e^{i\delta'}
                    & c_{12'}c_{23'}
                    & c_{12'}c_{13'}s_{23'}-s_{12'}s_{13'}e^{-i\delta'} \\
                -s_{13'}c_{23'}e^{i\delta'}
                    & -s_{23'}
                    & c_{13'}c_{23'}
                \end{pmatrix}$
            & $c_{12'}s_{12'}c_{13'}s_{13'}c_{23'}^2s_{23'}\sin(\delta')$
            \\
        \hline
        $U_{321}$
            & $\begin{pmatrix}
                c_{12'}c_{13'}
                    & s_{12'}c_{23'}-c_{12'}s_{13'}s_{23'}e^{-i\delta'}
                    & s_{12'}s_{23'}+c_{12'}s_{13'}c_{23'}e^{-i\delta'} \\
                -s_{12'}c_{13'}
                    & c_{12'}c_{23'}+s_{12'}s_{13'}s_{23'}e^{-i\delta'}
                    & c_{12'}s_{23'}-s_{12'}s_{13'}c_{23'}e^{-i\delta'} \\
                -s_{13'}e^{i\delta'}
                    & -c_{13'}s_{23'}
                    & c_{13'}c_{23'}
                \end{pmatrix}$
            & $c_{12'}s_{12'}c_{13'}^2s_{13'}c_{23'}s_{23'}\sin(\delta')$
            \\
        \hline
    \end{tabular}
    \label{tab:matJ}
\end{table}

\begin{table}
    \centering
    \caption{Formulas for the four parameters in a given parameterization of $U$ in terms of the absolute value of elements of $U$ and the Jarlskog. Since each mixing angle is in the interval $[0,90^\circ)$, $\cos(\theta_{ij}')=\sqrt{1-\sin[2](\theta_{ij}')}$. The formulas for the mixing angles are derived from the magnitudes of the ``simple" elements (see definition in the text) of $U$. The formulas for $\cos(\delta')$ are derived from the magnitude of a ``complicated" element of $U$. The formulas for $\sin(\delta')$ were derived from the Jarlskog.}
    \label{tab:parameters}
    \begin{tabular}{|l||c|c|c|c|c|}
        \hline
        $U_{ijk}$ & $\sin(\theta_{12}')$ & $\sin(\theta_{13}')$ & $\sin(\theta_{23}')$ & $\cos(\delta')$ & $\sin(\delta')$ \\
        \hline\hline
        $U_{123}$
            & $\frac{\abs{U_{e2}}}{\sqrt{1-\abs{U_{e3}}^2}}$
            & $\abs{U_{e3}}$
            & $\frac{\abs{U_{\mu3}}}{\sqrt{1-\abs{U_{e3}}^2}}$
            & $\frac{\abs{U_{\mu1}}^2\qty(1-\abs{U_{e3}}^2)^2-\abs{U_{e2}}^2\abs{U_{\tau3}}^2-\abs{U_{e1}}^2\abs{U_{e3}}^2\abs{U_{\mu3}}^2}{2\abs{U_{e1}}\abs{U_{e2}}\abs{U_{e3}}\abs{U_{\mu3}}\abs{U_{\tau3}}}$
            & $\frac{J\qty(1-\abs{U_{e3}}^2)}{\abs{U_{e1}}\abs{U_{e2}}\abs{U_{e3}}\abs{U_{\mu3}}\abs{U_{\tau3}}}$
            \\
        \hline
        $U_{132}$
            & $\abs{U_{e2}}$
            & $\frac{\abs{U_{e3}}}{\sqrt{1-\abs{U_{e2}}^2}}$
            & $\frac{\abs{U_{\tau 2}}}{\sqrt{1-\abs{U_{e2}}^2}}$
            & $\frac{\abs{U_{\mu1}}^2\qty(1-\abs{U_{e2}}^2)^2-\abs{U_{e3}}^2\abs{U_{\tau2}}^2-\abs{U_{e1}}^2\abs{U_{e2}}^2\abs{U_{\mu2}}^2}{2\abs{U_{e1}}\abs{U_{e2}}\abs{U_{e3}}\abs{U_{\mu2}}\abs{U_{\tau2}}}$
            & $\frac{J\qty(1-\abs{U_{e2}}^2)}{\abs{U_{e1}}\abs{U_{e2}}\abs{U_{e3}}\abs{U_{\mu2}}\abs{U_{\tau2}}}$
            \\
        \hline
        $U_{213}$
            & $\frac{\abs{U_{\mu1}}}{\sqrt{1-\abs{U_{\mu3}}^2}}$
            & $\frac{\abs{U_{e3}}}{\sqrt{1-\abs{U_{\mu3}}^2}}$
            & $\abs{U_{\mu3}}$
            & $\frac{\abs{U_{e1}}^2\qty(1-\abs{U_{\mu3}}^2)^2-\abs{U_{\mu2}}^2\abs{U_{\tau3}}^2-\abs{U_{e3}}^2\abs{U_{\mu1}}^2\abs{U_{\mu3}}^2}{2\abs{U_{e3}}\abs{U_{\mu1}}\abs{U_{\mu2}}\abs{U_{\mu3}}\abs{U_{\tau3}}}$
            & $\frac{J\qty(1-\abs{U_{\mu3}}^2)}{\abs{U_{e3}}\abs{U_{\mu1}}\abs{U_{\mu2}}\abs{U_{\mu3}}\abs{U_{\tau3}}}$
            \\
        \hline
        $U_{231}$
            & $\abs{U_{\mu1}}$
            & $\frac{\abs{U_{\tau1}}}{\sqrt{1-\abs{U_{\mu1}}^2}}$
            & $\frac{\abs{U_{\mu3}}}{\sqrt{1-\abs{U_{\mu1}}^2}}$
            & $\frac{\abs{U_{e3}}^2\qty(1-\abs{U_{\mu1}}^2)^2-\abs{U_{\mu2}}^2\abs{U_{\tau1}}^2-\abs{U_{e1}}^2\abs{U_{\mu1}}^2\abs{U_{\mu3}}^2}{2\abs{U_{e1}}\abs{U_{\mu1}}\abs{U_{\mu2}}\abs{U_{\mu3}}\abs{U_{\tau1}}}$
            & $\frac{J\qty(1-\abs{U_{\mu1}}^2)}{\abs{U_{e1}}\abs{U_{\mu1}}\abs{U_{\mu2}}\abs{U_{\mu3}}\abs{U_{\tau1}}}$
            \\
        \hline
        $U_{312}$
            & $\frac{\abs{U_{e2}}}{\sqrt{1-\abs{U_{\tau2}}^2}}$
            & $\frac{\abs{U_{\tau1}}}{\sqrt{1-\abs{U_{\tau2}}^2}}$
            & $\abs{U_{\tau2}}$
            & $\frac{\abs{U_{\mu1}}^2\qty(1-\abs{U_{\tau2}}^2)^2-\abs{U_{e2}}^2\abs{U_{\tau3}}^2-\abs{U_{\mu2}}^2\abs{U_{\tau1}}^2\abs{U_{\tau2}}^2}{2\abs{U_{e2}}\abs{U_{\mu2}}\abs{U_{\tau1}}\abs{U_{\tau2}}\abs{U_{\tau3}}}$
            & $\frac{J\qty(1-\abs{U_{\tau2}}^2)}{\abs{U_{e2}}\abs{U_{\mu2}}\abs{U_{\tau1}}\abs{U_{\tau2}}\abs{U_{\tau3}}}$
            \\
        \hline
        $U_{321}$
            & $\frac{\abs{U_{\mu1}}}{\sqrt{1-\abs{U_{\tau1}}^2}}$
            & $\abs{U_{\tau1}}$
            & $\frac{\abs{U_{\tau2}}}{\sqrt{1-\abs{U_{\tau1}}^2}}$
            & $\frac{\abs{U_{e3}}^2\qty(1-\abs{U_{\tau1}}^2)^2-\abs{U_{\mu1}}^2\abs{U_{\tau2}}^2-\abs{U_{e1}}^2\abs{U_{\tau1}}^2\abs{U_{\tau3}}^2}{2\abs{U_{e1}}\abs{U_{\mu1}}\abs{U_{\tau1}}\abs{U_{\tau2}}\abs{U_{\tau3}}}$
            & $\frac{J\qty(1-\abs{U_{\tau1}}^2)}{\abs{U_{e1}}\abs{U_{\mu1}}\abs{U_{\tau1}}\abs{U_{\tau2}}\abs{U_{\tau3}}}$
            \\
        \hline
    \end{tabular}
\end{table}
\end{landscape}

In the PDG parameterization, experiments have measured all of the mixing angles (though the measurements of $\theta_{23}$ are the least precise), but $\delta$ remains largely unconstrained.
Plans have been made to measure $\delta$ with good precision in future experiments \cite{Abe:2014oxa,Abi:2020evt,Baussan:2013zcy}, and this paper aims at discussing the meaning of the precision claimed by these measurements.
In the rest of this paper, the complex phase in $U_{\rm PDG}$ will be called $\dPDG$ while the complex phase in other parameterizations will be called $\delta'$.
Recently, T2K has placed some constraints on $\dPDG$ \cite{Abe:2018wpn,Abe:2019vii}\footnote{While NOvA also has sensitivity to $\dPDG$, its data is currently not constraining \cite{Acero:2019ksn}.}; for now, we will consider $\dPDG$ to be completely unconstrained and then, in section \ref{sec:discussion}, discuss the interplay between the T2K measurements and different parameterizations of the mixing matrix.

\section{\texorpdfstring{$\delta$}{delta} Comparison}
\label{sec:comparison}
\begin{figure}
\centering
\includegraphics[width=0.32\textwidth]{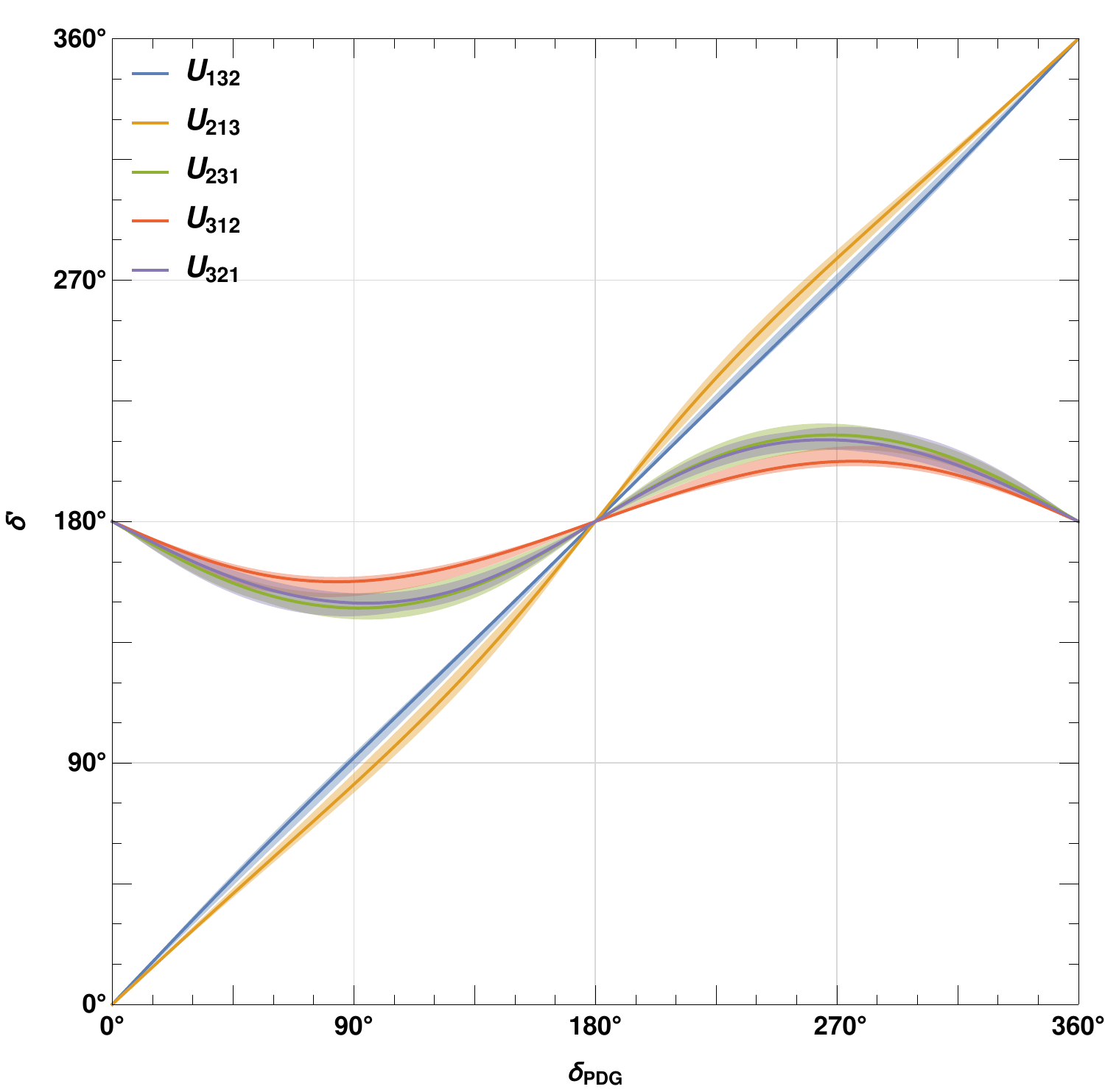}
\includegraphics[width=0.32\textwidth]{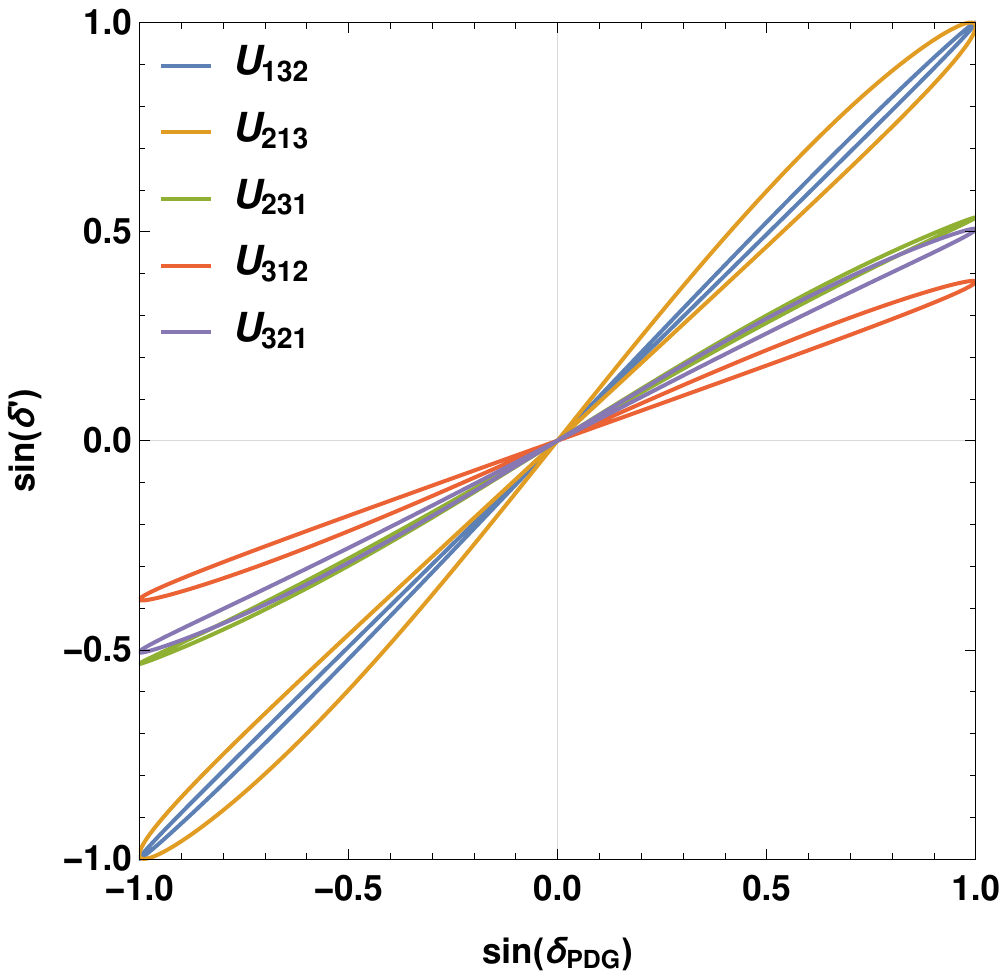}
\includegraphics[width=0.32\textwidth]{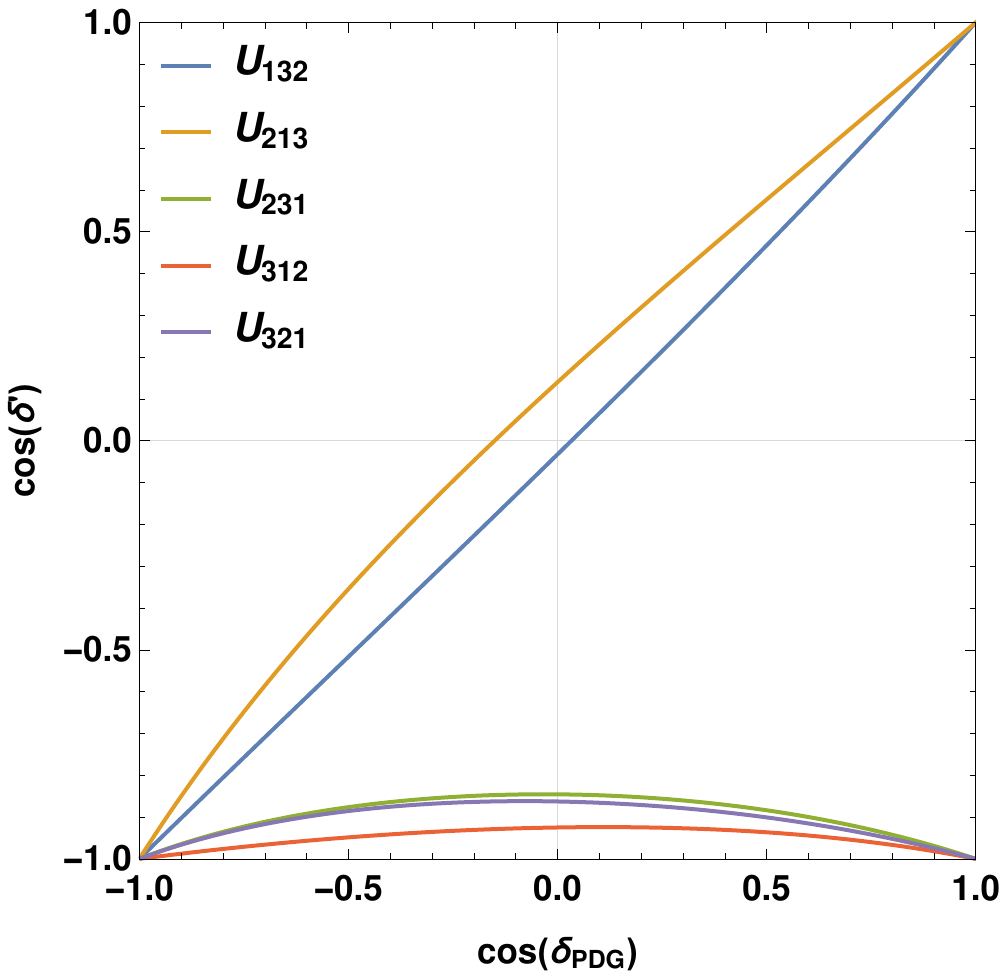}
\caption{Comparison of the CP-violating phase of $U_{\rm PDG}$ ($\delta_{\rm PDG}$) and that of other parameterizations of the neutrino mixing matrix ($\delta'$). The shaded regions on the left plot show the range of variations of the curves over the 3 $\sigma$ range of mixing angles from \cite{Esteban:2018azc}.}
\label{fig:dcpCompare}
\end{figure}

In order to compare the values of $\delta$ between the various parameterizations of $U$, we use the best fit values for the mixing angles from \cite{Esteban:2018azc} ($\theta_{12}=33.8^\circ$, $\theta_{13}=8.61^\circ$, and $\theta_{23}=49.7^\circ$) in $U_{\rm PDG}$, which gives us the following for the magnitudes of the elements in the neutrino mixing matrix:
\begin{equation}
\abs{U}=
\begin{pmatrix}
0.822 & 0.550 & 0.150 \\
\sqrt{0.138+0.068\cos(\dPDG)}
& \sqrt{0.293-0.068\cos(\dPDG)}
& 0.754 \\
\sqrt{0.186-0.068\cos(\dPDG)}
& \sqrt{0.405+0.068\cos(\dPDG)}
& 0.640
\end{pmatrix}
\,\text{.}
\label{eq:absU}
\end{equation}
The preferred value of the Jarlskog for this parameterization as a function of the complex phase is
\begin{equation}
    J=0.0334\sin(\dPDG)\,\text{.}
    \label{eq:Jbestfit}
\end{equation}
In table \ref{tab:parameters}, we show how to calculate the four parameters ($\theta_{12}'$, $\theta_{13}'$, $\theta_{23}'$, and $\delta'$) in six different parameterizations of $U$ in terms of the magnitudes of the elements of the matrix and the Jarlksog.
Using the expressions in this table combined with equations \ref{eq:absU} and \ref{eq:Jbestfit}, we have plotted the relationship between $\delta'$ vs $\dPDG$ (as well as $\sin(\delta)$ and $\cos(\delta)$) in fig.~\ref{fig:dcpCompare}.
Fig.~\ref{fig:dcpCompare} also shows that the effect of varying the mixing angles within their 3 $\sigma$ allowed values has a marginal impact on the values of $\delta'$ in other parameterizations. Varying the mixing angles significantly outside of these ranges, however, can have a dramatic effect on this figure.

We note that there are loops in the $\sin(\delta)$ plot but not in the $\cos(\delta)$ plot.
That is, $\cos(\dPDG)$ uniquely determines $\cos(\delta')$, but $\sin(\dPDG)$ does not uniquely determine $\sin(\delta')$.
This is because $\cos(\delta')$ comes from the norm of the elements of the matrix, and thus only depends on $\cos(\dPDG)$, while $\sin(\delta')$ comes from the Jarlskog, so it depends on $\sin(\dPDG)$ and $\cos(\dPDG)$ through the mixing angles.

An interesting feature of this plot is that for some of the alternative parameterizations of $U$, namely $U_{231}$, $U_{312}$, and $U_{321}$, $\delta'$ is restricted to a small domain just by our constraint of the three PDG mixing angles. The parameterizations for which $\delta'$ is bounded and whether it is bounded about $\delta'=0$ or $\delta'=180^\circ$ are dependent on the mixing angles.
Based on current measurements, the restriction is driven by the size of $\theta_{13}$ in the PDG parameterization or, alternatively, the relative smallness of $|U_{e3}|$.
One can see this by looking at $\abs{U_{e3}}$ in each parameterization, which is $s_{13}$ in $U_{\rm PDG}$ (see table \ref{tab:Ue3}).
In each of the parameterizations for which $\delta'$ is unbounded in fig.~\ref{fig:dcpCompare}, $\abs{U_{e3}}$ is simple, whereas in each of the parameterizations for which $\delta'$ is bounded, $\abs{U_{e3}}$ is complicated.
Thus, in order to get the comparatively small value of $|U_{e3}|$ in those parameterizations for which $|U_{e3}|=\sqrt{A+B\cos(\delta')}$ with $A$ and $B$ comparatively large, we must have $\cos(\delta')\sim-1$.
We can approximate the effect of small $|U_{e3}|$ on the allowed range of $\delta'$ in terms of PDG parameters by leveraging the smallness of $s_{13}$ to find
\begin{equation}
\max_{\dPDG}[\cos(\delta')]\approx \frac12 d_{ijk}^2-1\,,
\label{eq:approx cos deltap}
\end{equation}
with
\begin{align}
d_{231}\approx s_{13}\frac{1-s_{12}^2c_{23}^2}{s_{12}c_{12}s_{23}c_{23}} & =0.57\,, \qquad
d_{312}\approx s_{13}\frac{1-c_{12}^2s_{23}^2}{s_{12}c_{12}s_{23}c_{23}}=0.39\,, \nonumber\\
\text{and}\qquad
& d_{321}\approx s_{13}\frac{1-s_{12}^2s_{23}^2}{s_{12}c_{12}s_{23}c_{23}}=0.54
\label{eq:dijk}
\end{align}
for the three parameterizations with complicated $|U_{e3}|$.
We then see that $\sin(\delta')$ is approximately contained within $\pm d_{ijk}$.
In fact,
\begin{equation}
\sin(\delta')\approx d_{ijk}\sin(\dPDG)\,,
\label{eq:approx sind}
\end{equation}
provides a simple approximation for the expressions shown in fig.~\ref{fig:dcpCompare} and is accurate at the $\lesssim10\%$ level (eq.~\ref{eq:approx cos deltap} is accurate to $<2\%$).
From eq.~\ref{eq:dijk}, we can easily see how the three different parameterizations behave.
For example, we see that $U_{213}$ and $U_{321}$ should be quite similar since they differ only by $c_{23}^2\leftrightarrow s_{23}^2$, which are quite similar.
Thus, swapping $U_{231}$ and $U_{321}$ is the same (up to higher order corrections) as changing the octant.
Meanwhile, we see that the slope of $\sin(\delta')$ as a function of $\sin(\dPDG)$ is the smallest for $U_{312}$ (and thus $\delta'$ is the most constrained), since it has a factor $c_{12}^2$ instead of the factor of $s_{12}^2$.
This is all consistent with the top right panel in fig.~\ref{fig:dcpCompare}.

\begin{figure}
\centering
\includegraphics[width=0.32\textwidth]{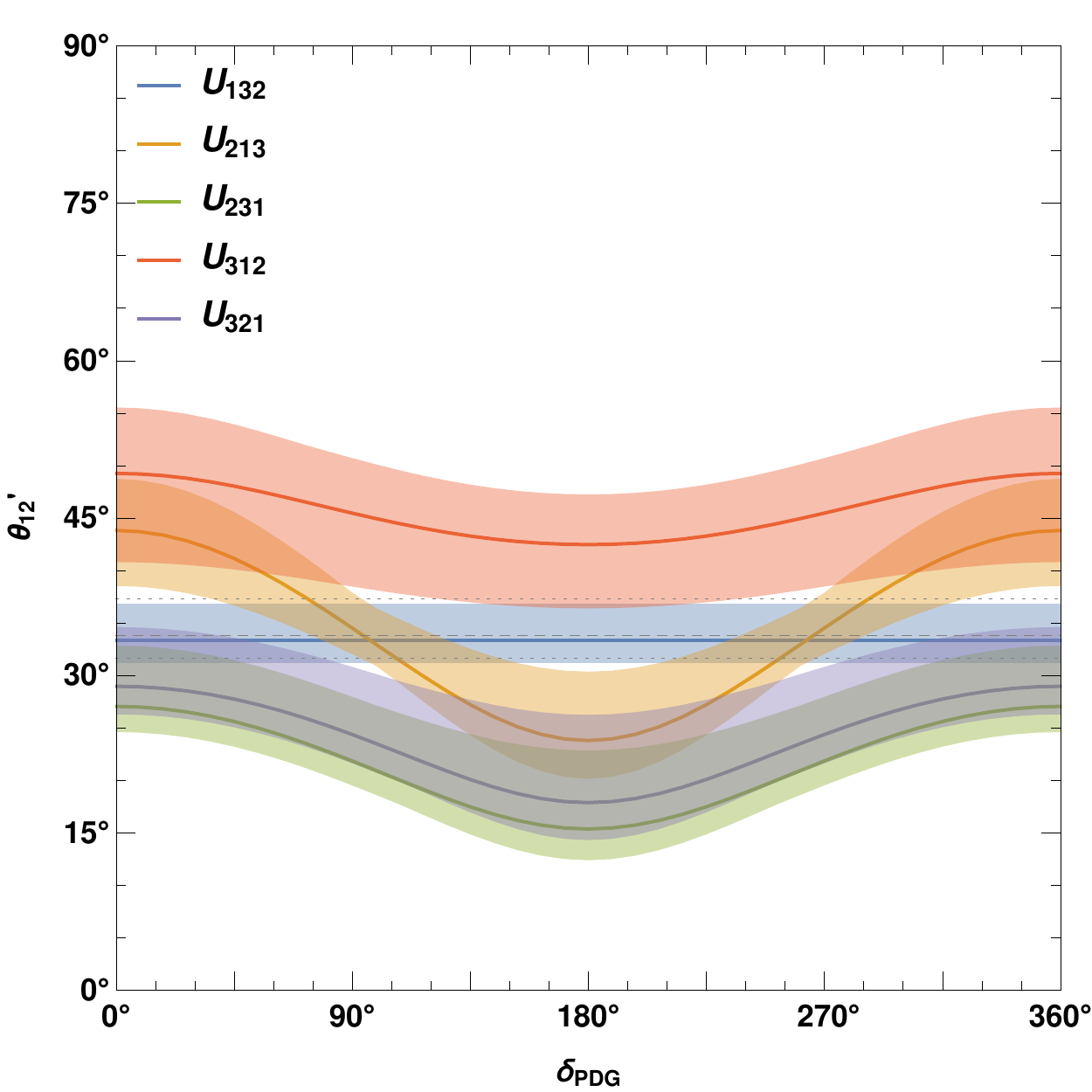}
\includegraphics[width=0.32\textwidth]{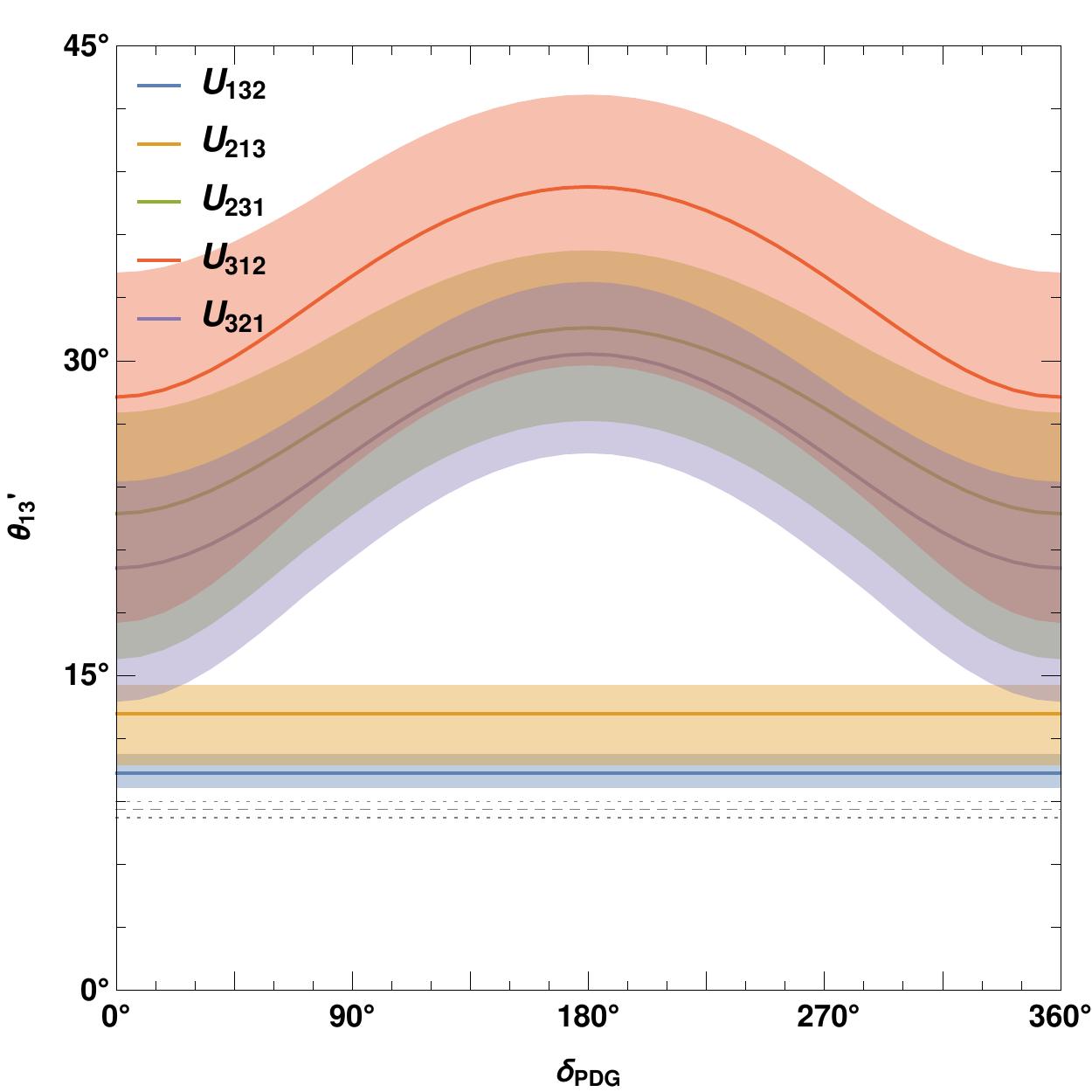}
\includegraphics[width=0.32\textwidth]{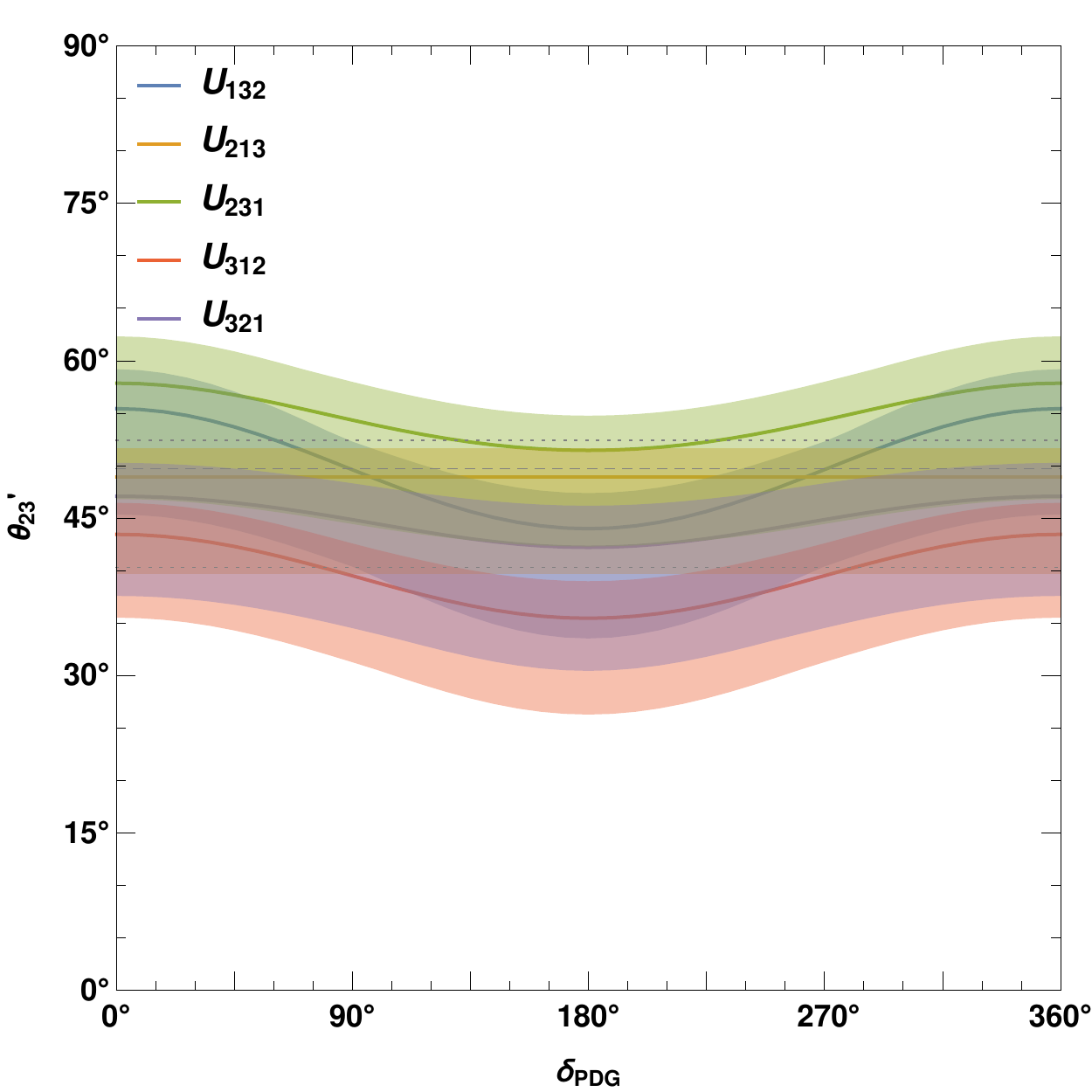}
\caption{Mixing angles in each parameterization as a function of $\dPDG$. The shaded regions show the range of variations of the curves over the 3 $\sigma$ range of mixing angles from \cite{Esteban:2018azc} while the dashed (dotted) gray lines are best fit values (maxima and minima of 3 $\sigma$ ranges) in the standard parameterization.}
\label{fig:mixingangles}
\end{figure}

We show the values of the three mixing angles in the different parameterizations in fig.~\ref{fig:mixingangles}, along with bands representing the region covered by these curves when the mixing angles in $U_\text{PDG}$ are allowed to vary independently within their 3 $\sigma$ ranges.
We note that $\theta_{12}'$ is only independent of $\dPDG$ in $U_{132}$ while $\theta_{23}'$ is only independent of $\dPDG$ in $U_{213}$.
On the other hand, $\theta_{13}'$ is independent of $\dPDG$ in both $U_{132}$ and $U_{213}$.
We can also see how the precision on the different oscillation parameters changes in different parameterizations\footnote{In the $U_{123}$ parameterization, there is some dependence on the constraints from $\dPDG$; these correlations are ignored here for simplicity.}.
Notably $\theta_{13}'$ is quite a bit less precise in other parameterizations, in particular those with complicated $U_{e3}$ elements.

\begin{figure}
\centering
\includegraphics[width=0.5\textwidth]{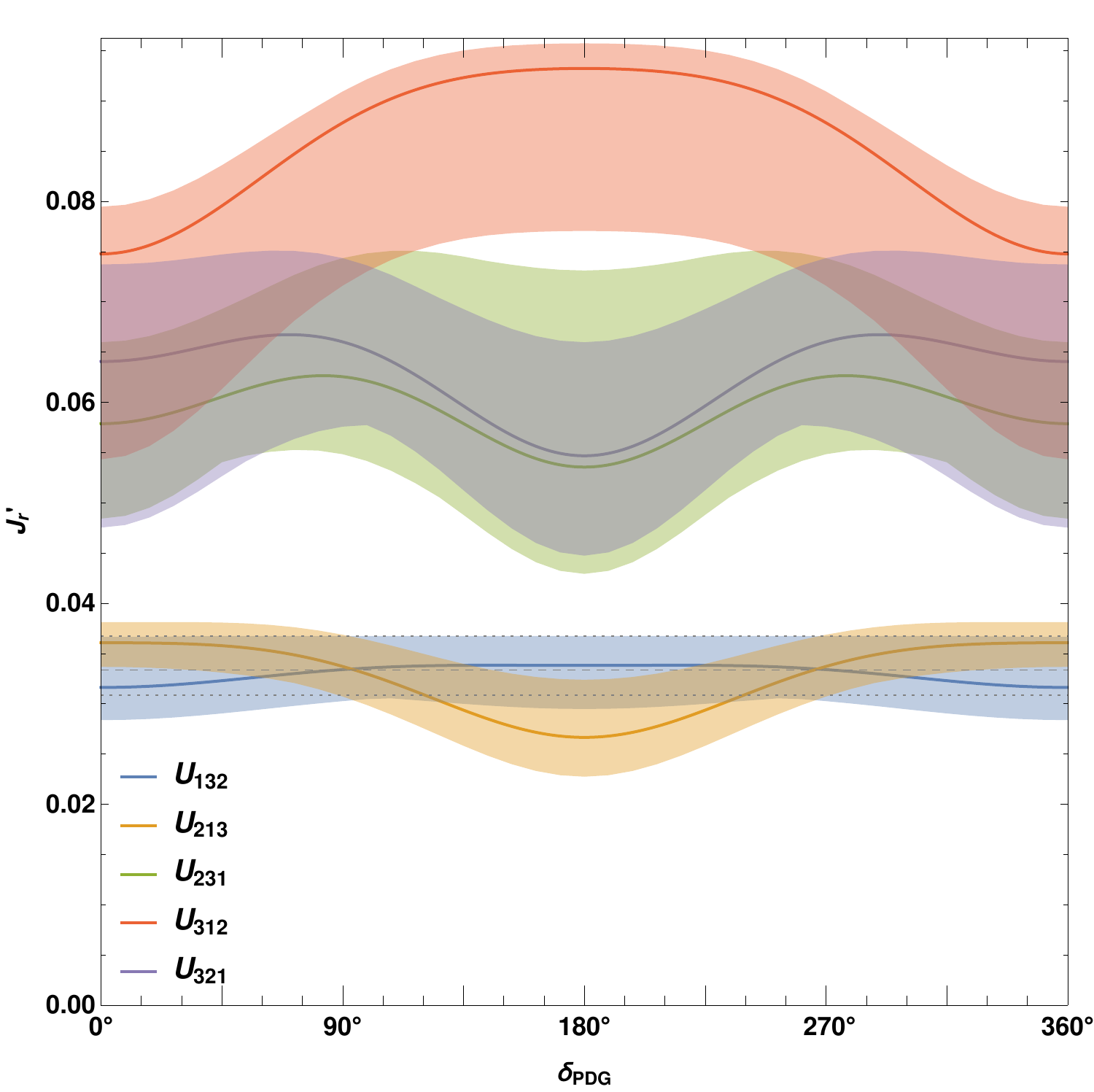}
\caption{The ``reduced Jarlskog" $J_r'\equiv\frac{J}{\sin(\delta')}$ for each parameterization as a function of $\dPDG$. The shaded regions show the range of variations of the curves over the 3 $\sigma$ range of mixing angles from \cite{Esteban:2018azc}. The dashed gray line is $J_r$ for $U_\text{PDG}$ using the best fit mixing angles from \cite{Esteban:2018azc} and the dotted gray lines are the maximum and minimum of $J_r$ for $U_\text{PDG}$ when allowing the mixing angles to vary over their 3 $\sigma$ ranges.
The sharp features are due to switching the preferred octant of $\theta_{23}$.}
\label{fig:Jr}
\end{figure}

We also show the reduced Jarlskog in different parameterizations as a function of $\dPDG$ in fig.~\ref{fig:Jr}.
The reduced Jarlskog is defined by,
\begin{equation}
J_r'\equiv\frac{J'}{\sin(\delta')}\,\text{.}
\end{equation}
(See table \ref{tab:matJ} for explicit expressions in each of the parameterizations considered.)
This quantity represents the contribution to the Jarlskog from the mixing angles in a given parameterization.
We can see that in the parameterizations with complicated expressions for $U_{e3}$ ($U_{231}$, $U_{312}$, and $U_{321}$), the reduced Jarlskog is notably much larger than it is in the parameterizations with simple expressions for $U_{e3}$ ($U_{123}$, $U_{132}$, and $U_{213}$).
This provides another means of showing why $\sin(\delta')$ in some parameterizations is restricted in its allowed range, as shown in fig.~\ref{fig:dcpCompare} and eqs.~\ref{eq:approx cos deltap}-\ref{eq:approx sind}.

\begin{landscape}
\begin{table}
    \centering
    \caption{$U_{e3}$ in various parameterizations of $U$. The primed variables ($\theta_{ij}'$, $s_{ij'}\equiv\sin(\theta_{ij}')$, $c_{ij'}\equiv\cos(\theta_{ij}')$, $\delta'$) denote the parameters in the alternative parameterization and the unprimed variables ($\theta_{ij}$, $\dPDG$, $c_\delta\equiv\cos(\dPDG)$) denote the parameters in $U_{\rm PDG}$. We also show an approximation for $|U_{e3}|$ in terms of the PDG parameterization variables at $\cos(\dPDG)\approx0$.}
    \label{tab:Ue3}
    \begin{tabular}{|c||c|c|c|}
        \hline
        $U_{ijk}$ 
            & $U_{e3}(\theta_{12}',\theta_{13}',\theta_{23}',\delta')$
            & $\abs{U_{e3}(\delta',\dPDG,\theta_{12}=33.8^\circ,\theta_{13}=8.61^\circ,\theta_{23}=49.7^\circ)}$
            & $\approx\abs{U_{e3}(\delta',c_\delta\approx 0,\theta_{12},\theta_{13},\theta_{23})}$
            \\
        \hline\hline
        $U_{132}$
            & $c_{12'}s_{13'}$
            & $0.150$
            & $s_{13}$
            \\
        \hline
        $U_{213}$
            & $s_{13'}c_{23'}$
            & $0.150$
            & $s_{13}$
            \\
        \hline
        $U_{231}$
            & $s_{13'}c_{23'}e^{-i\delta'}+s_{12'}c_{13'}s_{23'}$
            & $\frac{\sqrt{0.108-0.007c_\delta+0.005c_\delta^2+0.085\cos(\delta')\sqrt{1.62-0.17c_\delta-0.34c_\delta^2+0.07c_\delta^3}}}{0.861-0.068c_\delta}$
            & $\frac{2s_{12}c_{12}s_{23}c_{23}}{1-s_{12}^2c_{23}^2}\abs{\cos(\delta'/2)}$
            \\
        \hline
        $U_{312}$
            & $c_{12'}s_{13'}e^{-i\delta'}+s_{12'}c_{13'}s_{23'}$
            & $\frac{\sqrt{0.105-0.024c_\delta+0.005c_\delta^2+0.048\cos(\delta')\sqrt{4.74-2.04c_\delta-0.08c_\delta^2+0.07c_\delta^3}}}{0.595-0.068c_\delta}$
            & $\frac{2s_{12}c_{12}s_{23}c_{23}}{1-c_{12}^2s_{23}^2}\abs{\cos(\delta'/2)}$
            \\
        \hline
        $U_{321}$
            & $s_{12'}s_{23'}+c_{12'}s_{13'}c_{23'}e^{-i\delta'}$
            & $\frac{\sqrt{0.108+0.018c_\delta+0.005c_\delta^2+0.108\cos(\delta')\sqrt{1+0.29c_\delta-0.16c_\delta^2-0.03c_\delta^3}}}{0.814+0.068c_\delta}$
            & $\frac{2s_{12}c_{12}s_{23}c_{23}}{1-s_{12}^2s_{23}^2}\abs{\cos(\delta'/2)}$
            \\
        \hline
    \end{tabular}
\end{table}
\end{landscape}

\begin{figure}
\centering
\includegraphics[width=0.3\textwidth]{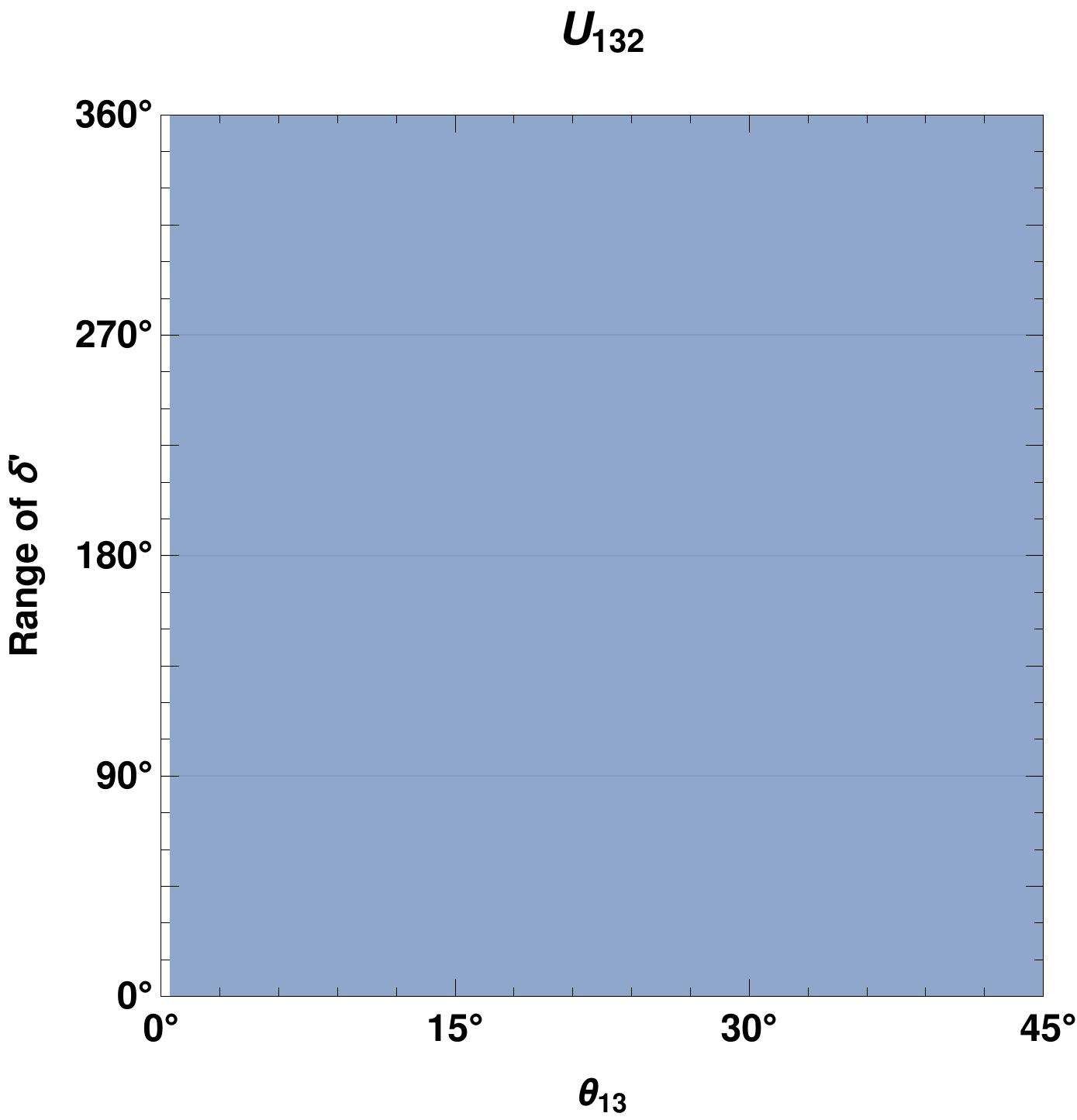}
\includegraphics[width=0.3\textwidth]{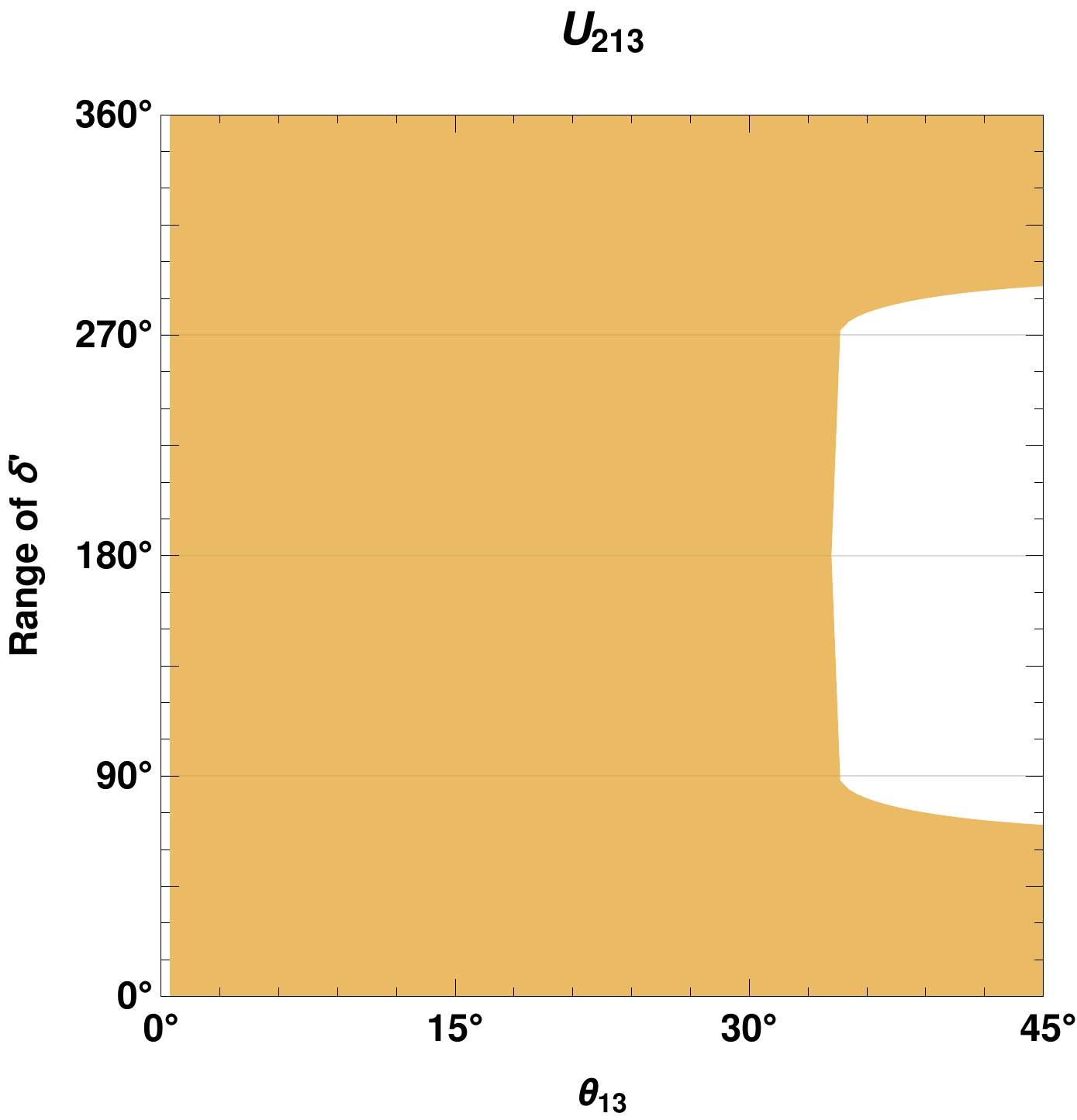}\\
\includegraphics[width=0.3\textwidth]{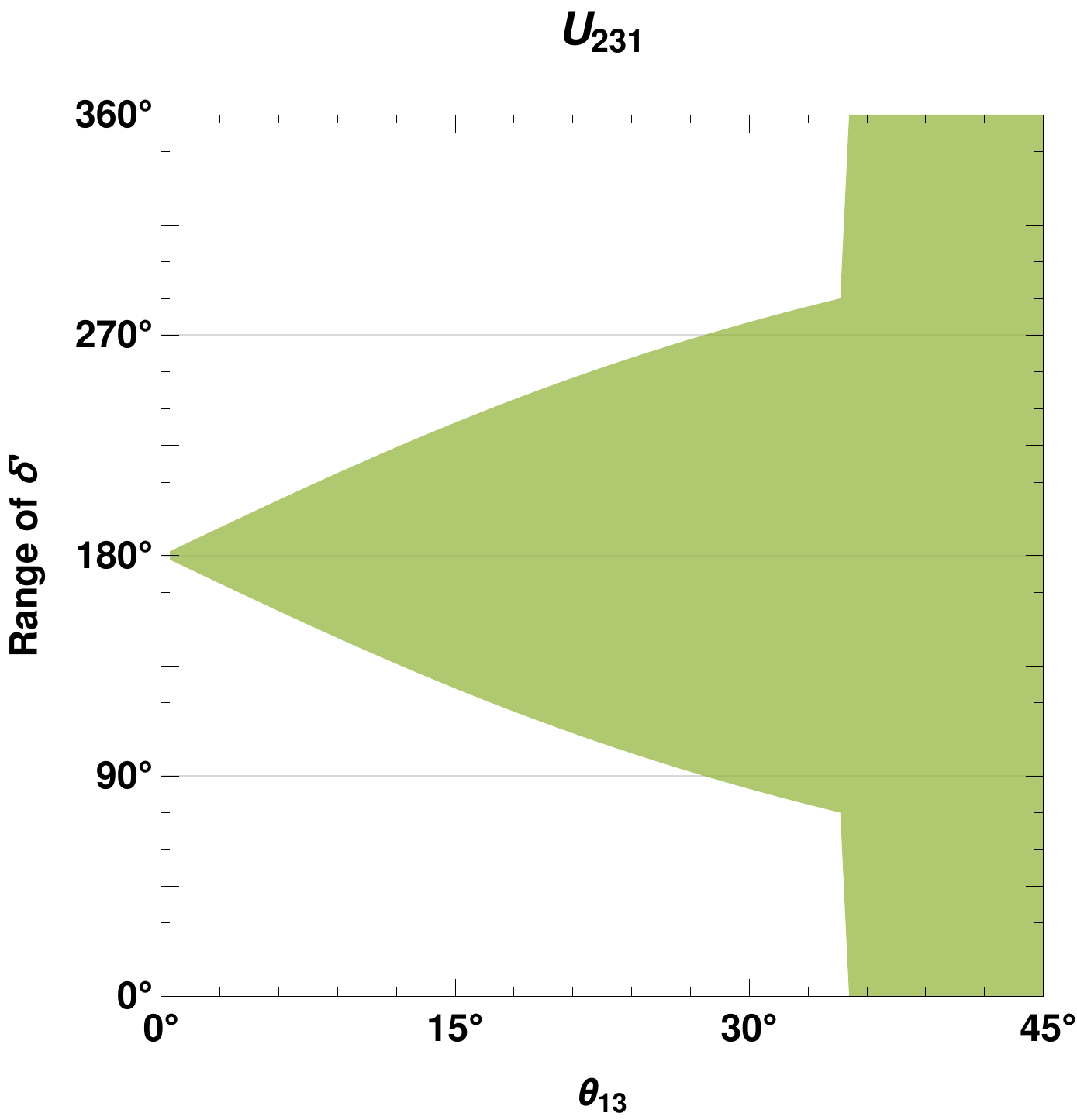}
\includegraphics[width=0.3\textwidth]{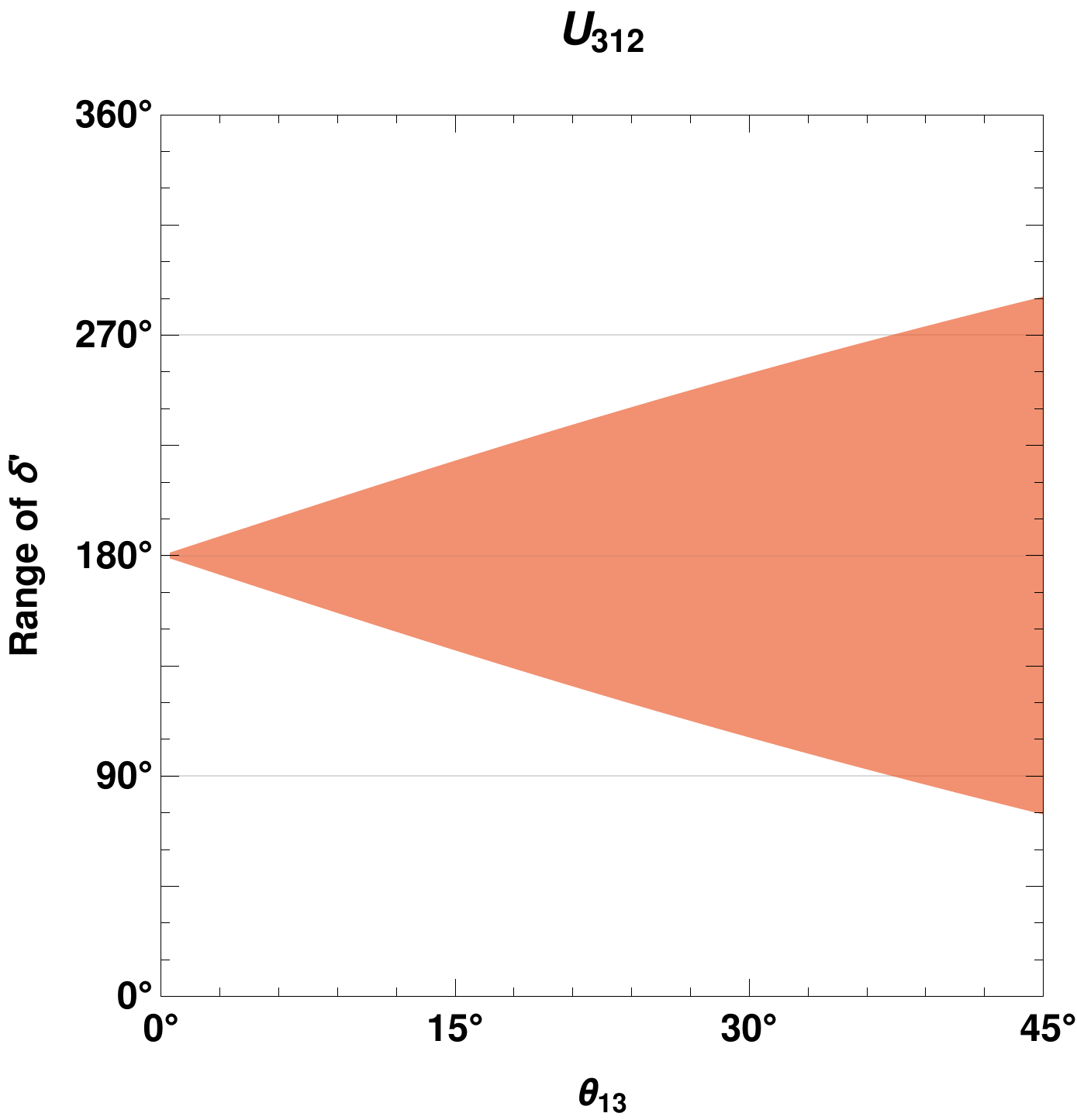}
\includegraphics[width=0.3\textwidth]{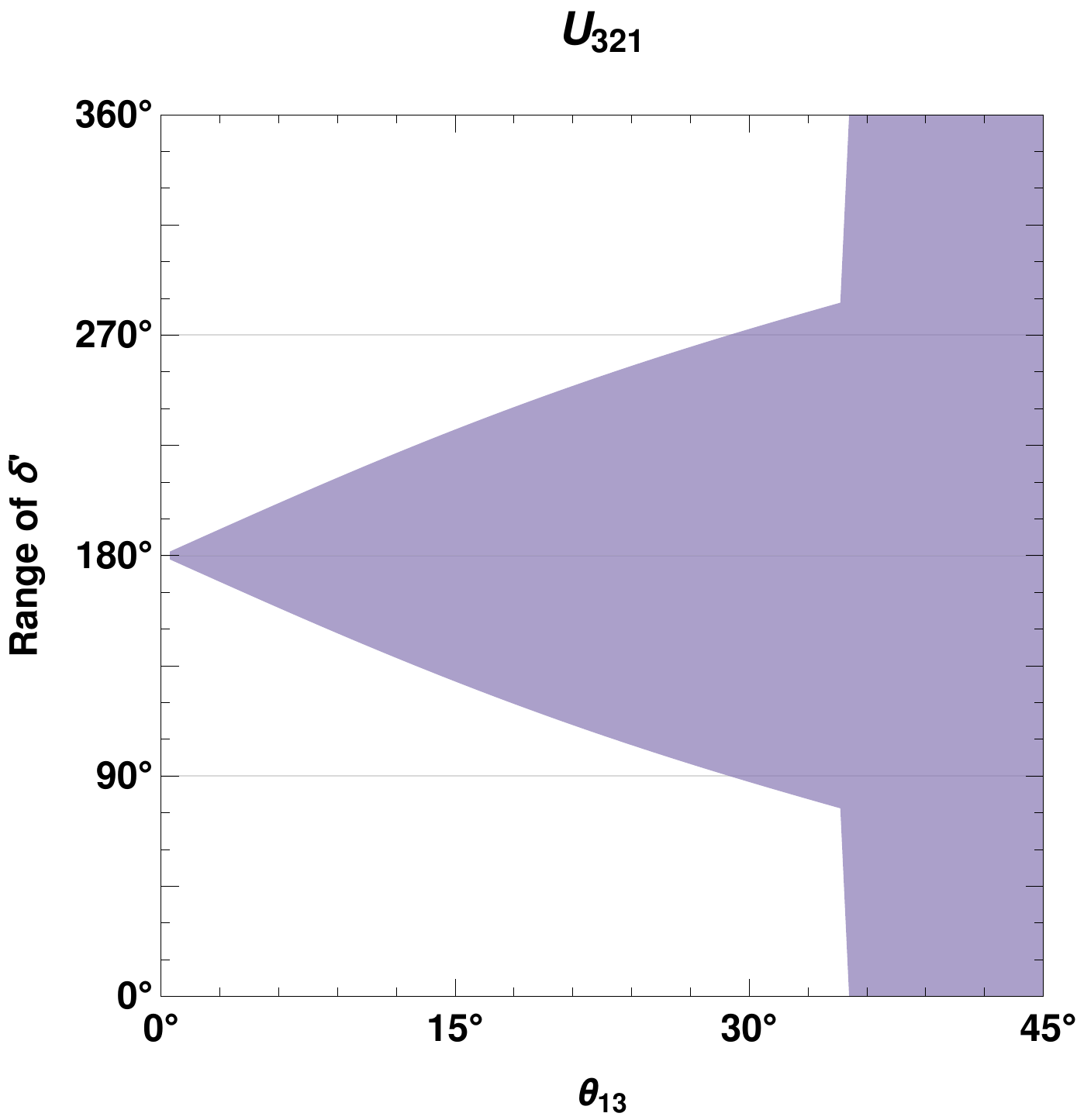}
\caption{The allowed range of $\delta'$ in different parameterizations as a function of $\theta_{13}$ in the PDG parameterization.
We have fixed the other mixing angles to their best fit values while $\delta_{\rm PDG}$ is unconstrained.}
\label{fig:allowed range}
\end{figure}

To further understand the impact of $|U_{e3}|$ on the tightly constrained nature of $\cos(\delta')$ in other parameterizations, we plotted the allowed range (assuming $\dPDG$ is unconstrained) of $\delta'$ for various different parameterizations as a function of $\theta_{13}$ in the PDG definition in fig.~\ref{fig:allowed range}, with all the other mixing angles fixed.
In the $U_{132}$ parameterization, $\delta'$ is always unconstrained for any value of $\theta_{13}$ up to 45$^\circ$.
For $U_{213}$, $\delta'$ is unconstrained until $\theta_{13}>35^\circ$ past which point $\cos(\delta')$ is constrained to be near 1.
In the three remaining parameterizations, we see that $\cos(\delta')$ is constrained to be near $-1$ and that the allowed region increases as $\theta_{13}$ increases roughly linearly with $\theta_{13}$ for smaller values of $\theta_{13}$, which is consistent with $d_{ijk}\propto s_{13}$ (see eq.~\ref{eq:dijk}).

\begin{figure}
\centering
\includegraphics[width=0.5\textwidth]{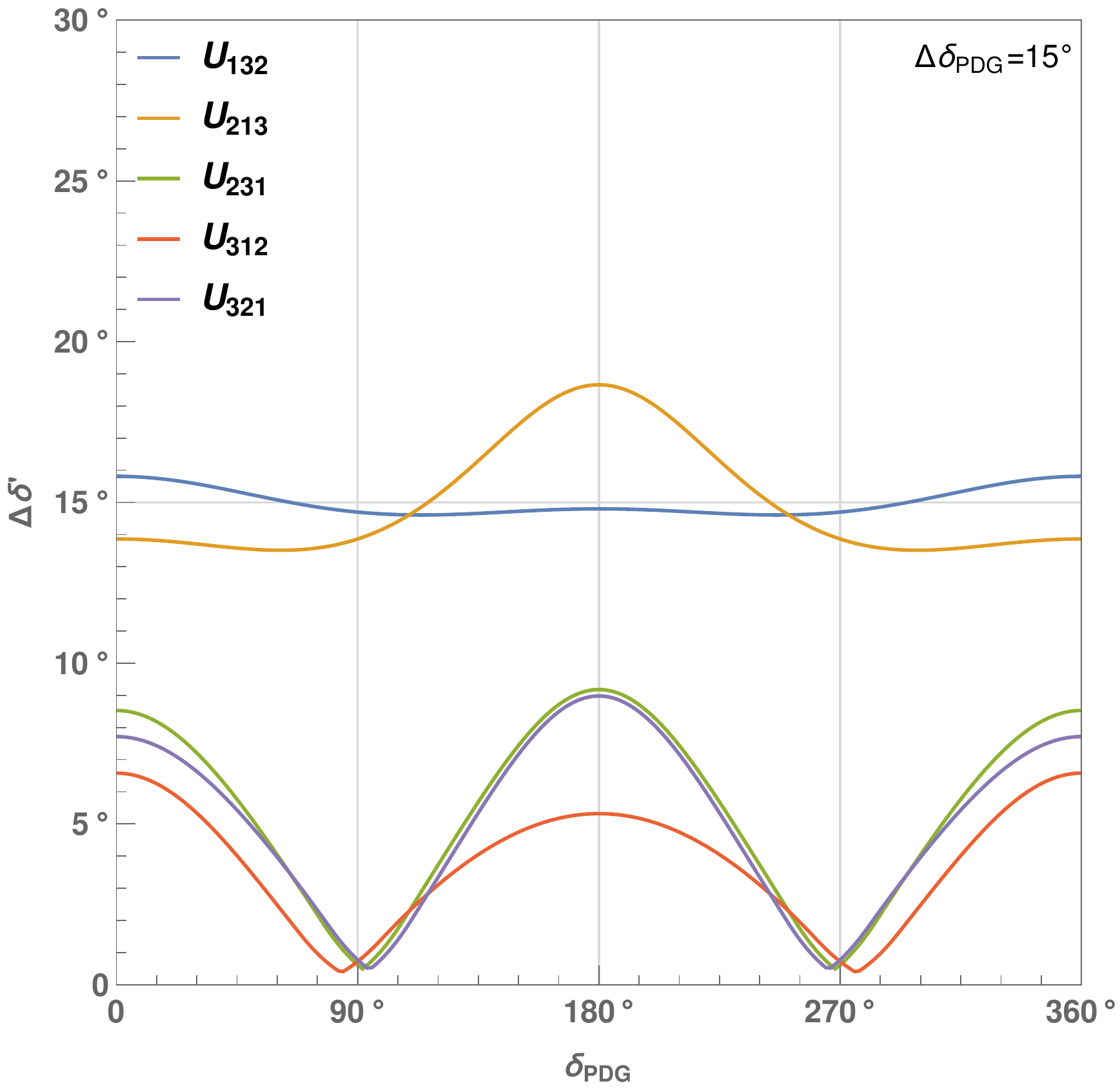}
\caption{Uncertainty of the CP-violating phase of alternate parameterizations of the neutrino mixing matrix ($\delta'$) as a function of $\delta_{\rm PDG}$, given an uncertainty on $\delta_{\rm PDG}$ of $15^\circ$.}
\label{fig:Deltadcp}
\end{figure}

Next generation long-baseline accelerator experiments are aiming to not only detect CPV, but measure it with some precision.
Various targets on the precision on $\dPDG$ have been presented based on theoretical concerns motivated by discrinimating among different flavor symmetries.
The 2013 Snowmass report stressed $\Delta\dPDG=15^\circ$ as an important goal \cite{deGouvea:2013onf} and a more recent theory overview emphasized $\Delta\dPDG=10^\circ$ \cite{Petcov:2018snn} as a useful target.
We take as a benchmark number $\Delta\dPDG=15^\circ$ and plot the corresponding $\Delta\delta'$ (with fixed mixing angles) in the other parameterizations for various values of $\dPDG$ (see fig.~\ref{fig:Deltadcp}).
We find that in different parameterizations of the lepton mixing matrix, a precision on $\dPDG$ of 15$^\circ$ will lead to extremely different precision in different parameterizations, in particular in the three parameterizations with complicated expressions for $U_{e3}$.
In fig.~\ref{fig:Deltadcp}, we see that while $U_{132}$ and $U_{213}$ aren't too different from the PDG parameterization (as seen before), in the other three parameterizations, a measurement of $\dPDG$ to within $15^\circ$ precision could result in $\sim1^\circ$ precision on $\delta'$, depending on the parameterization and the value of $\dPDG$.
In addition, since the Jarlskog (and thus the precision on the Jarlskog) is the same in each parameterization, fig.~\ref{fig:Deltadcp} indicates that the precision on the mixing angles, in particular $J_r\equiv c_{12}s_{12}c_{13}^2s_{13}c_{23}s_{23}$, is comparable in $U_{132}$ and $U_{213}$ as in the PDG parameterization.
Meanwhile, since $\delta'$ (and thus $\sin\delta'$) becomes more precise in $U_{231}$, $U_{312}$, and $U_{321}$, the mixing angles must become less precise in those parameterizations to compensate, as confirmed in fig.~\ref{fig:Jr}.
This is due to the fact that $U_{231}$, $U_{312}$, and $U_{321}$ have complicated $U_{e3}$ elements while the others all have simple $U_{e3}$ elements.

\begin{figure}
\centering
\includegraphics[width=0.6\textwidth]{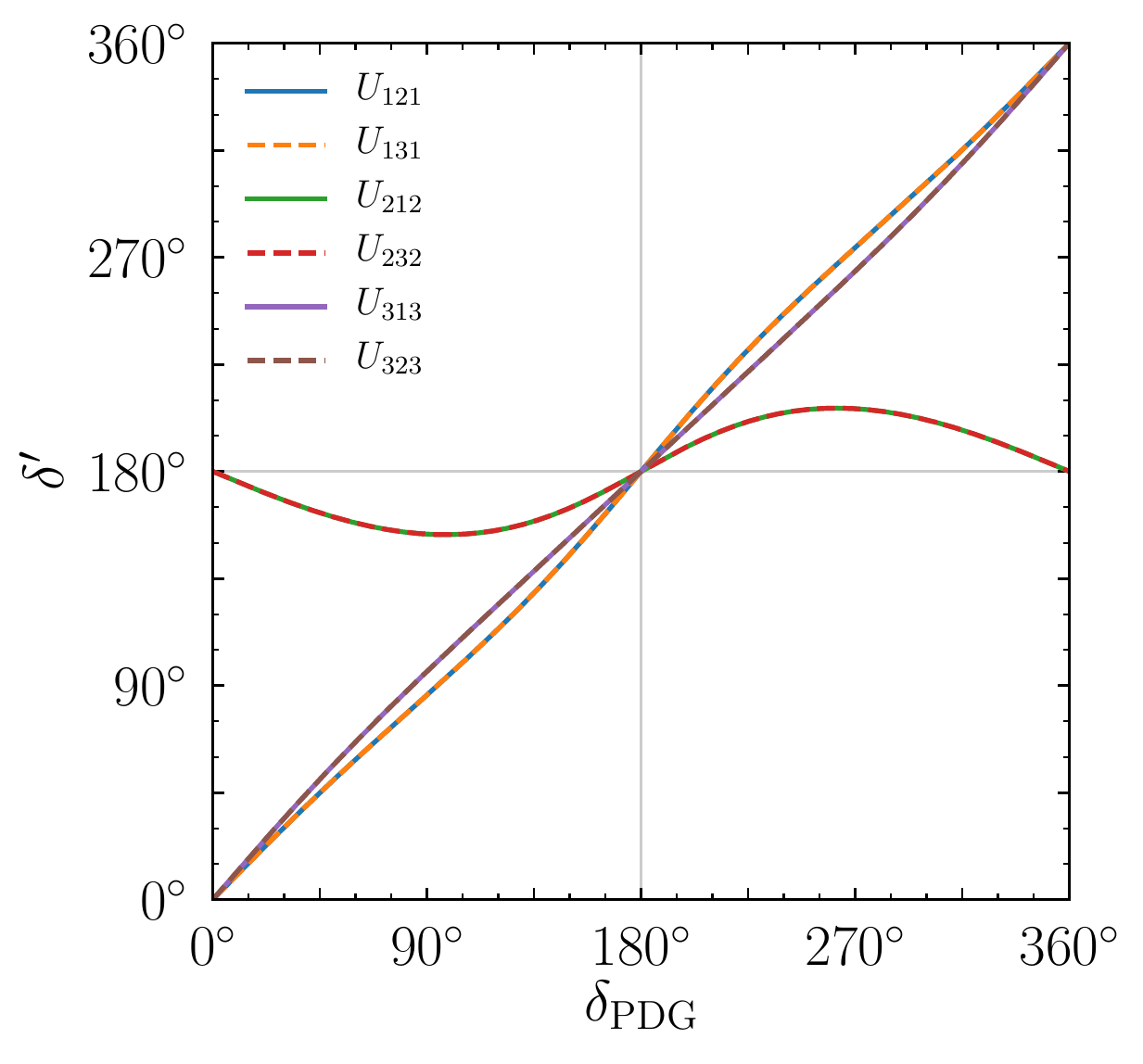}
\caption{The same as the left panel of fig.~\ref{fig:dcpCompare} except for parameterizations with a repeated rotation.
Note that $\delta'$ in the parameterizations $U_{iji}$ and $U_{iki}$ are equal.}
\label{fig:dcpCompareRepeated}
\end{figure}

In fig.~\ref{fig:dcpCompareRepeated} we also show $\delta'$ as a function of $\dPDG$ for six parameterizations involving a single repeated rotation axis\footnote{Note that the complex phase must be associated with either the first or the third rotation in the repeated rotations $U_{iji}$ parameterizations while it can be associated with any of the rotations in the $U_{ijk}$ parameterizations.} of the form $U_{iji}$.
We make two observations.
First, the relationships between $\delta'$ and $\dPDG$ are very similar in these parameterizations as they are in the $U_{ijk}$ style parameterizations.
In fact, they look nearly identical, but they have slight differences.
Second, we note that $\delta'$ in parameterizations of the forms $U_{iji}$ and $U_{iki}$ are equivalent for $i$, $j$, and $k$ all different.

\section{Discussion}
\label{sec:discussion}
As we have seen in fig.~\ref{fig:dcpCompare}, in some parameterizations, in particular $U_{213}$, $U_{312}$, and $U_{321}$, $\cos(\delta')$ is restricted to be $\lesssim-0.8$, and correspondingly, $|\sin(\delta')|\lesssim0.6$.
That is, if the lepton mixing matrix was parameterized differently, we would already know that $150^\circ\lesssim\delta'\lesssim210^\circ$ (and, given T2K's preference for $170^\circ\lesssim\dPDG\lesssim360^\circ$, this implies that $170^\circ\lesssim\delta'\lesssim210^\circ$).
This highlights the fact that $\dPDG=\pm90^\circ$ by itself is not maximal in any fundamental way; it does lead to the largest amount of CPV allowed \emph{given the other measured oscillation parameters}.

In much of the previous discussion and figures, we have assumed that we have measured the three oscillation parameters perfectly and have no information on $\dPDG$.
In reality, of course, each of the four parameters are somewhat constrained, although $\dPDG$ is largely unconstrained at the moment except via T2K appearance data.
We have verified, however, that even over the allowed 3 $\sigma$ range of $40^\circ\lesssim\theta_{23}\lesssim50^\circ$ (and even quite a bit beyond this), $\cos(\delta')$ is restricted in 3 of the parameterizations to a region near $-1$, and there is a corresponding constraint on $\sin(\delta')$, as shown by the shaded regions in fig.~\ref{fig:dcpCompare}.

The strongest constraint on CPV currently comes from T2K \cite{Abe:2019vii}.
T2K has measured a high amount of neutrino appearance and a low amount of anti-neutrino appearance, implying that $J<0$ at close to 3 $\sigma$.
In addition, they seem to see slightly more neutrino events and fewer anti-neutrino events than allowed by $\dPDG=270^\circ$ given other constraints, although this tension is at low significance.
Nonetheless, T2K has interpreted this as a constraint on $\theta_{13}$, which (very weakly) suggests a larger value than that from precise medium baseline reactor experiments \cite{Adey:2018zwh,Seo:2016uom,Abe:2014bwa}.
In this analysis, however, they assume priors on solar parameters.
If, instead, a prior on $\theta_{13}$ from reactors was used since it is extremely well measured, one would find that this slight tension could be more somewhat more easily accommodated in terms of a constraint on solar parameters.

We now take this slight tension seriously and consider the implications if it persists with additional data.
If it is confirmed at high precision, then we would look to the constraint on the other oscillation parameters, since the measurement depends on $c_{13}^2s_{13}c_{12}s_{12}c_{23}s_{23}\frac{\Delta m^2_{21}}{\Delta m^2_{31}}$ up to matter effects.
Based on the best fit value for $\theta_{13}$ from T2K, this results in a 14\% excess in this quantity.
The 1 $\sigma$ allowed regions for $c_{13}^2s_{13}$, $c_{12}s_{12}$, $\Delta m^2_{31}$, and $\Delta m^2_{21}$ are 1.4\%, 1.1\%, 1.3\%, and 2.8\%.
This suggests that the best place to understand any tension in the T2K data with the rest of the oscillation data is in terms of the solar mass splitting\footnote{While it is interesting to note that there was a slight tension in the solar mass splitting data, it goes in the wrong direction to explain the T2K tension.}.
We note that the matter effect only slightly modifies the situation at T2K, as the corrections to the Jarlskog can be determined in a straightforward fashion \cite{Denton:2019yiw}, as can the corrections to the $\Delta m^2_{ij}$'s \cite{Denton:2016wmg,Denton:2018hal}.

In addition, T2K's constraint on $\dPDG$ is that it is constrained to within $[165^\circ,358^\circ]$ at 3 $\sigma$, which is about half the allowed space using a uniform prior on $\dPDG$\footnote{The Haar measure applied to lepton mass mixing indicates that the correct prior on the complex phase is uniform in $\delta$ under the assumption of anarchy, instead of $\sin(\delta)$ or other possible choices \cite{deGouvea:2012ac}.}.
However, we find that if the lepton mixing matrix was parameterized as $U_{312}$, the allowed range on the complex phase is only $[173^\circ,208^\circ]$ including the 3 $\sigma$ uncertainty on the oscillation parameters.
This is only 7.5\% of the total $\delta'$ space, compared with 53.6\% of the total $\dPDG$ space.
Similarly, small regions exist for the $U_{231}$ and $U_{321}$ parameterizations.
Note that this means that in the $U_{312}$ parameterization, the uncertainty is already down to $\Delta\delta'=18^\circ$ at 3 $\sigma$ CL, accounting for the uncertainty in the oscillation parameters.

Thus, we propose that a more useful metric for quantifying CP violation is amount of allowed Jarlskog space.
Extracting the Jarlskog only requires (up to matter effects) knowing the ratio of $\Delta m^2$'s (and really just $\Delta m^2_{21}$, as T2K measures $\Delta m^2_{31}$) as opposed to $\sin(\dPDG)$, which also requires knowing $\theta_{13}$, $\theta_{12}$, and $\theta_{23}$.
The Jarlskog is always within the range $J\in[-\frac1{6\sqrt3},\frac1{6\sqrt3}]$, which numerically is $[-0.096,0.096]$.
These maximum (minimum) values of $J$ occur when $\theta_{12}=\theta_{23}=45^\circ$, $\theta_{13}=\atan(1/\sqrt2)=35.3^\circ$, and $\sin(\dPDG)=+1$ ($\sin(\dPDG)=-1$).
From nu-fit v4.0, it is found that the allowed range is reduced to $[-0.033,0.033]$ without including any $\sin(\dPDG)$ information.
This represents 35\% of the total allowed space, and it mostly comes from the fact that $\theta_{13}=8.6^\circ$ is small compared to the value for maximal CPV, $35.3^\circ$.
A subleading contribution comes from the fact that $\theta_{12}=34^\circ$, compared to $45^\circ$, which results in a reduction in the allowed space to 93\%.
Thus given T2K's measurement, the allowed space of the Jarlskog is about $22\%$, which is a parameterization independent statement.

We also note that certain parameterizations are very similar.
In particular, we see in fig.~\ref{fig:dcpCompare} that $U_{123}$ (the PDG parameterization) is quite similar to $U_{132}$.
In addition, $U_{231}$ is quite similar to $U_{321}$.
Thus, we can conclude that $U_2$ and $U_3$ roughly commute within the context of the discussion of CPV.
That is, the effect of $\delta'$ in the different parameterizations doesn't change much when commuting $U_2$ and $U_3$.
This makes sense since the associated angles in the PDG parameterization, $\theta_{13}$ and $\theta_{12}$ are the two smallest angles, so while rotations are not commutative in general, the correction scales with the size of the rotations.

Before concluding this section, we make one comment on the usefulness of $\delta$ in a given parameterization (or $\dPDG$ in the usual parameterization).
While we have extensively demonstrated the shortcomings of $\delta$ in general and as a proxy for understanding CP violation, as well as the benefit of using the Jarlskog to quantify the amount of CP violation in the leptonic mass matrix, there is still some value in the $\delta$ quantity.
In particular, it comes from the goal of physics to measure everything.
That is, measuring the Jarlskog with arbitrary precision, as well as three rotation angles, leaves one sign undetermined: the sign of $\cos\delta$.
While the sign of $\cos\delta$ does not affect CP violation, it is still a fundamental parameter in the Standard Model and is physical, and thus it must be measured.
Thus we do not advocate entirely ignoring measurements of $\delta$ as measuring parameters in the Standard Model is the primary goal of particle physics, but we do encourage experimentalists and theorists to focus on the Jarlskog when discussing CP violation.

\section{Conclusion}
\label{sec:conclusion}
Three of the four degrees of freedom in the lepton mixing matrix have been measured reasonably well.
In the standard PDG parameterization, these parameters are labeled $\theta_{12}$, $\theta_{13}$, and $\theta_{23}$.
The final parameter, $\dPDG$, is related to the amount of CP violation (CPV) in the lepton sector affecting neutrino oscillations.
While it is certainly the case that $\sin(\dPDG)=0$ implies no CPV in neutrino oscillations and $\sin(\dPDG)\neq0$ is proof of CPV (given that all three mixing angles are non-zero), $\dPDG$ and CPV in neutrino oscillations are not exactly equivalent concepts.
One clear way to see this is that in different parameterizations with one complex phase, the allowed range of the new phase $\delta'$ may be already severely constrained, as shown in fig.~\ref{fig:dcpCompare}, without even considering any constraint on $\dPDG$ coming from T2K or other experiments.
This also highlights the fact that the precision with which $\dPDG$ is measured is not truly fundamental (see fig.~\ref{fig:Deltadcp}), so the precision on the Jarlskog should be considered instead.
The cause of the surprisingly strong constraint on $\delta'$ in certain other parameterizations without any information on $\dPDG$ is the smallness of $|U_{e3}|$, which requires a fairly strong cancellation, which only happens for $\cos(\delta')$ near $-1$.

We use this time to revisit the topic of optimality with regards to the choice of parameterization for the leptonic mixing matrix in the context of neutrino oscillations, giving the presently known data\footnote{There are other potentially interesting criteria such as computational efficiency, for which $SU(N)$ generators may be more efficient than rotations \cite{Delgado:2014kpa}.}.
While this is of course a matter of taste, we here list some desirable features of a matrix:
\begin{itemize}
\item Since $|U_{e3}|$ is small, it is favorable to have a simple element there; this picks out parameterizations $U_{123}$, $U_{132}$, $U_{213}$, $U_{121}$, $U_{131}$, $U_{313}$, or $U_{323}$.
\item It is useful to be able to write approximate two-flavor oscillations as a simple function of a single mixing angle for the various experimental probes that are available to us.
By looking at solar ($U_{e2}$), medium-baseline reactor ($U_{e3}$), and long-baseline accelerator/atmospheric disappearance ($U_{\mu3}$), we find that the preferred parameterizations from table \ref{tab:parameters} (i.e. parameterizations in which the element concerned is simple) are:
\begin{center}
\def\y{\textcolor{ForestGreen}{\text{\ding{51}}}}
\def\n{\textcolor{red}{\text{\ding{55}}}}
\begin{tabular}{c|c|c|c|c|c|c}
&$U_{123}$&$U_{132}$&$U_{213}$&$U_{231}$&$U_{312}$&$U_{321}$\\\hline
$|U_{e2}|$  &\y&\y&\n&\n&\y&\n\\
$|U_{e3}|$  &\y&\y&\y&\n&\n&\n\\
$|U_{\mu3}|$&\y&\n&\y&\y&\n&\n
\end{tabular}\\[0.1in]
\begin{tabular}{c|c|c|c|c|c|c}
&$U_{121}$&$U_{131}$&$U_{212}$&$U_{232}$&$U_{313}$&$U_{323}$\\\hline
$|U_{e2}|$  &\y&\y&\y&\y&\n&\n\\
$|U_{e3}|$  &\y&\y&\n&\n&\y&\y\\
$|U_{\mu3}|$&\n&\n&\y&\y&\y&\y
\end{tabular}
\end{center}

\end{itemize}
The only parameterization that satisfies all of these conditions is $U_{123}$, which is the same as the PDG parameterization (without the Majorana phases).
The only remaining choice is the location of the oscillation phase on one of the three rotations: $U_1$, $U_2$, or $U_3$.
As this does not lead to any redefinition of the phase other than possibly a sign, its location is less important than the order of rotations.
Nonetheless, there are some theoretical benefits to putting the phase on $U_1$ (the $\theta_{23}$ rotation) instead of the more conventional $U_2$ (the $\theta_{13}$ rotation) when discussing the matter effect, see e.g.~\cite{Toshev:1991ku,Denton:2016wmg}.
For consistency, however, we recommend that in most applications, the conventional PDG parameterization, including associating the phase with the $\theta_{13}$ rotation, should be used; see eq.~\ref{eq:UPDG}.

In addition, one could examine the impact the choice of parameterization has on a given symmetry structure such tribimaximal, trimaximal, golden ratio, and others along with their modifications \cite{Vissani:1997pa,Barger:1998ta,Harrison:2002er,Xing:2002sw,He:2003rm,Albright:2008rp,Altarelli:2010gt}.
Since these are typically written in the PDG parameterization, the statements in this article apply in a straightforward fashion to each model.
That is, if a given model predicts a certain range for $\dPDG$, then in a different parameterization, that range would be given by fig.~\ref{fig:dcpCompare}.
Nonetheless, one could investigate whether or not the sum rules connected with various models exhibited any useful simplifications in certain parameterizations; see appendix \ref{sec:flavor models} for an example.

Finally, as neutrino oscillation measurements improve and the values of the lepton mixing matrix fall into sharp relief, we hope that when quantifying the amount of CPV present, the Jarlskog is used, which depends on all four parameters in any parameterization, instead of just $\dPDG$.

\begin{acknowledgments}
We thank Stephen Parke for useful conversations.
PBD acknowledges the United States Department of Energy under Grant Contract desc0012704.
The work presented here that RP did was supported by the U.S.~Department of Energy, Office of Science, Office of Workforce Development for Teachers and Scientists, Office of Science Graduate Student Research (SCGSR) program. The SCGSR program is administered by the Oak Ridge Institute for Science and Education (ORISE) for the DOE. ORISE is managed by ORAU under contract number DE-SC0014664. All opinions expressed in this paper are the authors' and do not necessarily reflect the policies and views of DOE, ORAU, or ORISE.
\end{acknowledgments}

\appendix

\section{Unitarity Violation}
\label{sec:uv}
The discussion in the main text has focused on the standard three flavor oscillation picture, but as testing new physics scenarios in neutrino oscillations is an important goal of the neutrino program \cite{Arguelles:2019xgp}, we here investigate the impact of new physics on the above discussion.
One popular general new physics scenario that directly modifies the mixing matrix is unitarity violation (UV) \cite{Antusch:2006vwa}.
In the various different UV schemes the accessible $3\times3$ mixing matrix is not unitary, likely because it the submatrix of a larger unitary matrix.
This is the expectation if neutrinos get their masses from one of the seesaw mechanisms \cite{Minkowski:1977sc,Mohapatra:1979ia,Schechter:1980gr} or due to any number of other new physics scenarios.
While not confirmed, there are several hints of unitary violation in the neutrino sector known as the Galium anomaly \cite{Giunti:2010zu,Kostensalo:2019vmv,Kostensalo:2020hbc}, the reactor anti-neutrino anomaly \cite{Mention:2011rk,Berryman:2019hme}, and LSND and MiniBooNE anomalies \cite{Aguilar:2001ty,Aguilar-Arevalo:2018gpe,Aguilar-Arevalo:2020nvw}.
In addition, there are some anomalies related to UV in the quark sector as well \cite{Tanabashi:2018oca,Belfatto:2019swo,Belfatto:2021jhf}, see appendix \ref{sec:ckm} for more on the CKM matrix, in general.

A common parameterization for UV is via the $\alpha$ matrix wherein the non-unitary mixing matrix, $N$, is presumably mostly\footnote{Note that this parameterization is not perturbative; it completely covers the parameter space of unitary violation.} unitary and thus is the product of a unitary matrix and another matrix describing the violation from unitarity,
\begin{equation}
N\equiv(1-\alpha)U\,,
\label{eq:N}
\end{equation}
where $U$ is a unitary matrix parameterized as one chooses with four parameters, and $\alpha$ is a $3\times3$ matrix,
\newcommand{\aee}{\alpha_{ee}}
\newcommand{\ame}{\alpha_{\mu e}}
\newcommand{\amm}{\alpha_{\mu\mu}}
\newcommand{\ate}{\alpha_{\tau e}}
\newcommand{\atm}{\alpha_{\tau\mu}}
\newcommand{\att}{\alpha_{\tau\tau}}
\begin{equation}
\alpha\equiv
\begin{pmatrix}
\aee&0&0\\
\ame&\amm&0\\
\ate&\atm&\att
\end{pmatrix}\,,
\label{eq:alpha}
\end{equation}
with real elements on the diagonal and complex elements off the diagonal. Thus the off-diagonal elements introduce three additional complex phases.
Note that in the literature there are variations on eqs.~\ref{eq:N}-\ref{eq:alpha} wherein the $\alpha$ matrix is upper triangular, or there is no $(1-)$ term such that the diagonal elements are $\sim1$, or the $\eta$ parameterization where all elements are populated and the matrix is Hermitian.
The UV parameters are all trivially related to each other without any effect on the unitary part of the matrix \cite{Blennow:2016jkn}, so we can fix ourselves to this parameterization without loss of generality.

Given a UV scheme, one can then evaluate how the parameters (in the unitary part or in the UV part) change as one changes the parameterization of the unitary matrix.
To do so, we performed the same numerical exercise as in fig.~\ref{fig:dcpCompare}.
We found that the presence of UV is completely decoupled from the issue of parameterizations of the unitary part of the matrix.
In particular, we found that for \emph{any} values of the parameters in the $\alpha$ matrix, large or small, real or complex, uniform or structured, the mapping between $\dPDG$ and $\delta'$ in the various different parameterizations is unaffected by the presence of UV.
This means that in the presence of UV the complex phase associated with the unitary part of the matrix can be completely distinguished from the rest of the matrix, although with realistic measurements this may or may not be the case.

\section{Quark Mixing}
\label{sec:ckm}
For completeness, we briefly comment on the same issue for quarks.
Quark mixing is governed by the CKM \cite{Cabibbo:1963yz,Kobayashi:1973fv} matrix $V$ and is often parameterized in the same way as the usual PDG parameterization for leptons, the $V_{123}$ parameterization.
In this parameterization, the four parameters are well measured to be \cite{Tanabashi:2018oca}
\begin{equation}
\theta_{12}=13.09^\circ\,,\quad
\theta_{13}=0.2068^\circ\,,\quad
\theta_{23}=2.323^\circ\,,\quad
\dPDG=68.53^\circ\,.
\end{equation}
It is interesting that, in this parameterization, $\dPDG$ is relatively close to 90$^\circ$ and thus $\sin\dPDG=0.93$ is close to one.
Yet it is well known that the total amount of CP violation in the quark sector is quite small as the Jarlskog coefficient is $J_{\rm CKM}/J_{\max}=3.1\e{-4}$ while $J_{\rm PMNS}/J_{\max}$ could be as large as 0.35, more than one thousand times larger, subject to upcoming measurements.

To further emphasize that $\delta$ is a parameterization dependent quantity, we point out that if the quark matrix is parameterized as $V_{212}$ we find that $\delta'=178.9^\circ$ leading to $\sin\delta'=0.0197$ which is nearly zero.
While these two choices of parameterizations describe the same underlying physics, if one only considers $\delta$ as an indication of CP violation, one will come to very different conclusions.

\section{Flavor Models}
\label{sec:flavor models}
We also investigate the impact of one's choice of parameterization on flavor models.
As an example, we consider the model \cite{Smirnov:2018luj,Smirnov:2019msn} wherein $U_{\rm PMNS}\simeq V_{\rm CKM}^\dagger U_{\rm BM}$ where $V_{\rm CKM}$ is the quark mixing matrix and $U_{\rm BM}$ is the bimaximal matrix \cite{Vissani:1997pa,Barger:1998ta}.
If one parameterizes the quark and bimaximal matrices as
\begin{equation}
V_{\rm CKM}\simeq
\begin{pmatrix}
c_C&s_Ce^{-i\phi}&0\\
-s_Ce^{i\phi}&c_C&0\\
0&0&1
\end{pmatrix}\,,\quad
U_{\rm BM}=
\begin{pmatrix}
\frac1{\sqrt2}&\frac1{\sqrt2}&0\\
-\frac12&\frac12&-\frac1{\sqrt2}\\
-\frac12&\frac12&\frac1{\sqrt2}
\end{pmatrix}\,,
\end{equation}
where $\theta_C$ is approximately the Cabibbo angle and $\phi$ is related to the complex phase in the quark matrix\footnote{In order to actually have CP violation in the quark matrix the other two quark mixing angles must be non-zero, but their impact on the leptonic parameters is small.}.

One then finds that
\begin{gather}
|U_{e2}|^2=\frac{s_C^2}4+\frac{c_C^2}2-\frac{s_Cc_C}{\sqrt2}c_\phi\,,\label{eq:Ue2}\\
J=-\frac1{4\sqrt2}s_Cc_Cs_\phi\,.\label{eq:J model}
\end{gather}
Thus eq.~\ref{eq:Ue2} determines $\phi$ and then eq.~\ref{eq:J model} determines $J$, subject to experimental uncertainties and higher order effects such as renormalization running.
Then using $J=J_r'\sin\delta'$, we can determine $\delta'$ in this model in various different parameterizations.
Using $|U_{e2}|=0.55$ and $s_C=0.225$, we find $c_\phi<-1$ which presumably implies that $J=0$ and thus $\delta'=0,\pi$ in all parameterizations.
We instead consider the more interesting case where $|U_{e2}|=0.60$.
Then we find that $c_\phi=0.78$ and $J=-0.024$ which leads to $\sin\dPDG=-0.69$.
This leads to predictions for $\delta'$ that vary by $34^\circ$ depending on one's choice of parameterization, as shown in table \ref{tab:model}.

While this example is specific to one particular model, the general conclusions remain the same.
Flavor models often provide predictions that can be interpreted as a specific value (or range of values) of $\dPDG$ in light of the fact that the other three parameters in the mixing matrix are relatively well measured now.
In a different parameterization the value and the range would change as shown in figs.~\ref{fig:dcpCompare} and \ref{fig:Deltadcp}.

\begin{table}
\centering
\caption{The prediction for $\delta'$ in different parameterizations from the model described in the text and for a slightly larger value of $|U_{e2}|$.}
\begin{tabular}{c|c}
Parameterization&$\delta'$ ($^\circ$)\\\hline
$U_{123}$&224\\
$U_{132}$&223\\
$U_{213}$&229\\
$U_{231}$&201\\
$U_{312}$&195\\
$U_{321}$&201\\\hline
$U_{121}$&228\\
$U_{131}$&228\\
$U_{212}$&200\\
$U_{232}$&200\\
$U_{313}$&222\\
$U_{323}$&222
\end{tabular}
\label{tab:model}
\end{table}

\bibliographystyle{JHEP}
\bibliography{delta}

\end{document}